%% file: pug.tex
\documentclass[11pt,a4paper]{article}
\pdfoutput=1
\usepackage{a4wide,lmodern,graphicx,verbatim,authblk}
\input{defs.tex}

\title{Interface tracking using patches}

\author[1,*]{Dag Lindbo}
\author[1]{Anna-Karin Tornberg} 
\affil[1]{Numerical Analysis, Royal Inst. of Tech. (KTH), 100 44
  Stockholm, Sweden}

\date{June 28, 2010}

\begin{document}
\maketitle

\thispagestyle{empty}
\let\oldthefootnote\thefootnote
\renewcommand{\thefootnote}{\fnsymbol{footnote}}
\footnotetext[1]{To whom correspondence should be addressed. Email:
  dag@kth.se }
\let\thefootnote\oldthefootnote

  \begin{abstract}
    A new method for interface tracking is presented. The interface
    representation, based on domain decomposition, provides the
    interface location explicitly, yet is Eulerian. This allows for
    well established finite difference methods on uniform grids to be
    used for the numerics of advecting the interface and other
    computations. CFL-stable and second order accurate explicit and
    implicit time-stepping methods are derived. Numerical results are
    given to substantiate stated convergence properties, as well as
    convergence in interface curvature and mass conservation. Our
    method is applied to a boundary integral formulation for Stokes
    flow, and the resulting integrals are analyzed and treated
    numerically to second order accuracy. Finally, we embed our method
    in the familiar immersed boundary- and immersed interface methods
    for two-phase Navier-Stokes flow.
  \end{abstract}

\section{Introduction}
Interface tracking has received considerable attention in the last 20
years. Much of this has been motivated by applications in fluid
mechanics, e.g. multiphase flows and crystal growth. There is also
strong interest in the mathematical community on a more general class
of problems that involve studying PDEs with dynamic free boundaries,
so called free boundary problems. The present work provides an
approach to interface tracking that differs at a basic level from the
established methods.

An interface may be thought of as a boundary between two or more
sub-domains -- in $\R^2$ represented as one or more curves, and in
$\R^3$ represented as surfaces. In the language of flow applications,
we let $\U(\cdot,t)$ denote a convective field that deforms the
interface according to the ODE
\begin{align}
  \dd{\iface}{t} = \U(\iface,t), \label{eq:interface_ode}
\end{align}
where $\iface$ denotes the interface, e.g. given in 2D by a parametric
curve, $\iface = \{(x(s,t), y(s,t)): t>0, s\in[a\,\, b) \}$. The task
of any interface tracking method is to provide a concrete
representation of $\iface$ and an equation of time evolution
equivalent to \eqref{eq:interface_ode} that lends itself to practical
computation.

There are several approaches present that provide such a
representation, e.g. implicit methods (including \emph{level set},
\emph{phase field} and \emph{volume of fluid} methods) and pure
Lagrangian methods (such as \emph{front tracking}). These are
typically introduced in the context of a particular application
domain, where additional questions often arise. We shall not attempt
to survey this entire field.

The level set method, introduced by Osher and Sethian in
\cite{Osher1988}, and with subsequent work in
e.g. \cite{Adalsteinsson1995, Sethian1999, Sussman1994, Olsson2005},
is widely used. This is a fundamentally Eulerian method where the
interface is not explicitly specified. Instead it is defined as a
contour, or level set, of a function which is defined over all of the
embedding space. This provides several advantages, including that it
generalizes well from 2D to 3D, and that topology changes can occur
without special treatment (though this feature is hard to control
since it depends directly on the discretization resolution). Notable
drawbacks include the fact that the interface tracking problem is now
discretized in one dimension higher than the interface itself (at a
computational cost), and that a sharp (explicit) interface
representation may be desired in various applications (including
e.g. surfactant problems \cite{Khatri2009, Khatri2011} and various
methods for enforcing interface jump conditions, such as the
\emph{immersed interface method} \cite{Lee2003, Li2006,
  Le2006}). Other implicit methods, such as phase field
\cite{Warren1995, Karma1996} and volume of fluid (VOF) methods
\cite{Hirt1981, Puckett1997} are widely used in computational
mechanics.

The front tracking methods introduced by Unverdi and Tryggvason in
\cite{Unverdi1992}, on the other hand, discretize
\eqref{eq:interface_ode} directly as a discrete set of points which
are evolved individually by integrating the ODE. Subsequent work can
be found in e.g. \cite{Tryggvason2001, Glimm1998}. Front tracking
methods preserve the dimensionality of the interface tracking problem
and provide the interface location sharply. An explicit
parametrization of the interface may none the less need to be
constructed (e.g. as in Glimm et. al \cite{Glimm1998}) to deal with
various issues, such as point distribution or computation of interface
properties, e.g. curvature.

Recently so called \emph{hybrid} methods have been introduced that
apply ideas from level set, front tracking and VOF methods to each
other in various permutations. See
e.g. \cite{Enright2002,Gaudlitz2008,Wang2009} and the references
therein. Whereas these methods have successfully remedied some flaws
of the basic methods, they clearly add practical and theoretical
complexity to already complicated methods.

In this paper we take a fundamentally different view on the
representation of the interface and its dynamics that provides a
method that is \emph{both Eulerian and explicit}. In essence, we
decompose the interface into a collection of segments, each of which
is described as a single valued function. Our inspiration for this
comes partly from two established fields: domain decomposition (see
e.g. Demmel \cite[Sec. 6.10]{Demmel1997}, and Toselli and Widlund
\cite{Toselli2004}) and differentiable manifolds (see e.g. Nakahara
\cite{Nakahara2003}). This work is a reformulation and extension of
the \emph{segment projection} method by Tornberg and Engquist
\cite{Tornberg2003, Engquist2002}. 

In addition to the conceptual distinctiveness of this approach, which
we think is relevant in the somewhat mature field of interface
tracking, our method provides some new possibilities that will help
tackle complex applications. Surfactant problems (i.e. where
additional PDEs are solved \emph{on the interface}) is one such
example, and contact line problems (i.e. where the interface interacts
with a solid wall in the presence of some flow) is another. For the
latter case it becomes possible with our method to pose boundary
conditions on the interface itself in a natural way. We also note that
this method is efficient and practical to implement, at least in
2D. In 3D the efficiency improvement over other methods is noteworthy,
but technical challenges with representing general closed surfaces
have prevented us from deploying our method in 3D in full
generality.

In Section \ref{sec:method} our method is introduced and the relevant
equations are derived. Then, in Section
\ref{sec:segment_discretization}, we present an explicit and implicit
numerical method for our equations, including the domain decomposition
approach and various related components. We give numerical results for
the interface tracing method in Section \ref{sec:num_res} and conclude
with two applications: a boundary integral method for interface Stokes
flow (Section \ref{sec:BIE_Stokes}), and immersed interface- and
boundary methods for two-phase Navier-Stokes flow (Section
\ref{sec:NS}).

\section{Interface tracking method in 2D} \label{sec:method} 

Let a simple plane curve be represented by $\iface \subseteq
\R^2$. Introduce a parametrization, $s \in [a\,\,b)$, of $\iface$ at
time $t$, $\iface = \iface(s,t) = (x(s,t),\,\, y(s,t))$, as in the
previous section. We will assume that the interface has sufficient
regularity, e.g. with $x$ and $y \in C^2$. The evolution of $\iface$
is described by the ODE \eqref{eq:interface_ode}. We will view the the
interface as set of \emph{segments}, $\segglob_i \subseteq \iface$,
$\cup \segglob_i = \iface $. Each such segment we will choose such
that it can be expressed as a single valued function, as presented in
Figure \ref{fig:interface_mark_seg}.
\begin{figure}
  \begin{minipage}[b]{0.45\linewidth}
    \centering
    \includegraphics{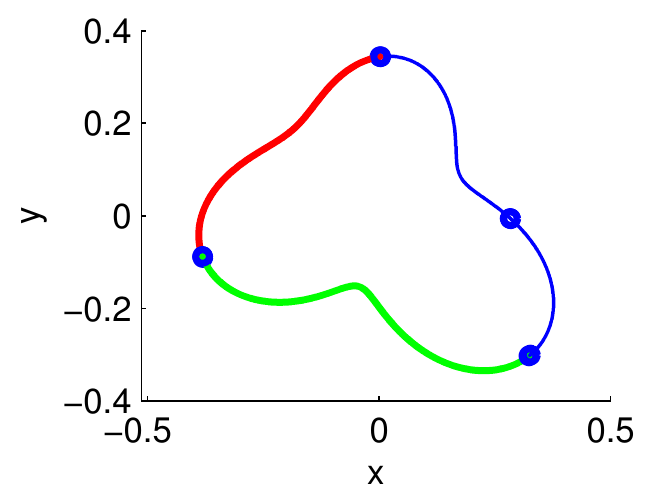}
  \end{minipage}
  \hspace{0.1cm}
  \begin{minipage}[b]{0.45\linewidth}
    \centering 
    \includegraphics{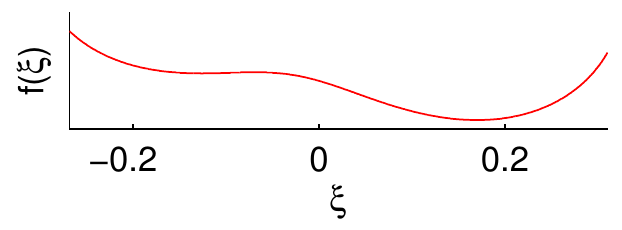}
    \includegraphics{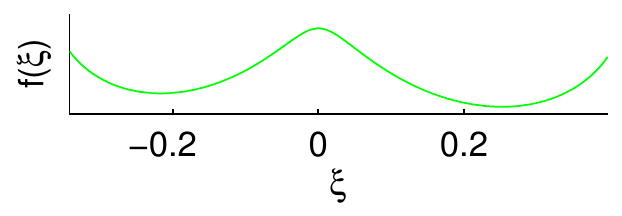}
  \end{minipage}
  \caption{Interface split into segments (ends marked with
    $\circ$), with two segments drawn bold, and their local
    representations.}
  \label{fig:interface_mark_seg}
\end{figure}
That is, assume there exists a in interval of $s' \subseteq s$ and
mapping $\map$ such that
\begin{align}
  \map \segglob = (\xi, f(\xi))
\end{align}
with $\segglob := \iface(s')$. We call this representation, $\map
\segglob := \segloc$, the segment in \emph{local} coordinates, as
opposed to $\segglob$, which we will refer to the segment in
\emph{global} coordinates. The mapping will be composed of a
translation, $\b$, and a unitary rotation, $\rot$, of some angle
$\theta$. We define this mapping and its inverse as
\begin{align}
  \map \x := \rot(\theta)\x + \b \label{eq:map}\\
  \imap \x := \map^{-1} \x = \rot(-\theta)(\x-\b), \label{eq:imap}
\end{align}
for any $\x \in \R^2$. The inverse mapping $\imap$ will naturally play
the role of transforming the segment from local to global coordinates,
i.e.
\begin{align*}
  \imap:\,\, \segloc \sra \segglob.
\end{align*}
Now we proceed to providing an equation for the time evolution of each
segment. Differentiate the transformation \eqref{eq:map} to obtain
\begin{align*}
  \dd{}{t}\map\segglob = 
  \begin{bmatrix}
    \left(\dd{}{t}\map\segglob\right)_1\\
    \left(\dd{}{t}\map\segglob\right)_2
  \end{bmatrix} =
  \begin{bmatrix}
    \dd{\xi}{t}\\
    \pd{f}{t} + \dd{\xi}{t} \pd{f}{\xi}
  \end{bmatrix}.
\end{align*}
Hence, we have
\begin{align*}
  \pd{f}{t} + \left(\dd{}{t}\map\segglob\right)_1 \pd{f}{\xi} = 
  \left(\dd{}{t}\map\segglob\right)_2\\
\end{align*}
which we may write more conveniently as
\begin{align}
  \pd{f}{t} + \left(\dd{}{t}\map\segglob\right) \cdot
  \left[\pd{f}{\xi},\quad -1\right] = 0. \label{eq:segment_pde}
\end{align}
We shall consider first the case where the mapping parameters $\theta$
and $\b$ are both constant and then, in Section \ref{sec:vf_decomp} we
present an extension where we have them as time dependent functions,
$\theta(t)$ and $\b(t)$.

For the first case it is evident that 
\begin{align*}
  \dd{}{t}\map\segglob = & \jac(\map \segglob ) \dd{\segglob}{t} \\
  = & \rot(\theta) \U(\segglob,t),
\end{align*}
where $\jac$ is the Jacobian and we have used the interface ODE
\eqref{eq:interface_ode}. Then we have the segment advection PDE given
by
\begin{align}
  \pd{f}{t} + v(\segglob,t) \pd{f}{\xi} = w(\segglob,
  t), \label{eq:segment_pde_static}
\end{align}
where we have introduced $\rot(\theta)\U(\segglob,t) =: (v, w)$. This is
an advection-type PDE with a source term. Note that the parameters $v$
and $w$ depend on the interface location, so the PDE is non-linear. In
Section \ref{sec:segment_discretization} we give an explicit and an
implicit method to treat this PDE numerically in a simple manner.

\subsection{Segment discretization} \label{sec:segment_discretization}

We now present numerical methods for treating the segment advection
equations \eqref{eq:segment_pde_static}. There are two primary issues
to address. First, as previously noted, we deal with non-linear
equations. However, these are hyperbolic -- in quasi-linear form --
and the non-linearity presents no major concern.

The second complication concerns boundary conditions and is more
interesting: \emph{there are no boundary conditions at the ends of
  each segment end if the interface is closed}. We have decomposed the
interface into segments, which leaves us with the requirement that
segments match where they meet and that smoothness is preserved
there. It is then natural that we devise a domain decomposition method
that solves the segment advection equation concurrently in time for
all segments.

We discretize the segment functions, $f(\xi)$ on uniform grids in
$\xi$, e.g. $\xi_i = a + i\dxi$. This constitutes a simplification
over other methods -- we can use difference formulas over 1D uniform
grids for our numerical treatment of the segment advection PDE as well
as all invariant interface quantities that we may need (such as
curvature, normals, quadrature formulas etc.).

The convective field, $\U(\cdot,t)$, may be an explicit function, as
in our test problems later in this paper. We also give examples later
where $\U$ is a functional over the interface (in a boundary integral
method), and, finally, a numerical solution operator to the
incompressible Navier Stokes equations. From the point of view of the
interface tracking method, $\U(\cdot,t)$ is a black box that orders a
convective field to the grid function on $\xi$.

In anticipation of these more general coefficient operators it is
sensible to introduce Strang splitting for the pair $(f,\U)$: Let $\U$
be discretized on staggered time levels, i.e. that 
\begin{equation}
  \label{eq:strang}
  \begin{split}
    \U^{n-1/2} \rightarrow \U^{n+1/2} \text{ using } f^n \text{ and}\\
    f^n \rightarrow f^{n+1} \text{ using } \U^{n+1/2}.    
  \end{split}
\end{equation}
It is clear for linear systems that this method maintains second order
accuracy, provided second order accurate time step schemes for $f$ and
$\U$ (if $\U$ is governed by a differential equation).

\subsection{Lax-Wendroff method for segments} \label{sec:laxw}
For the explicit method we will derive a Lax-Wendroff method with
central difference approximations in space (see e.g. the textbook by
LeVeque \cite{LeVeque2002}). The same method was used by Tornberg
et. al. in the segment projection method \cite{Tornberg2003}, and we
present it here for completeness. Consider
\begin{equation} \label{pde_coeff_gamma}
  f_t + v(\x(t),t) f_\xi = w(\x(t),t) 
\end{equation}
and let $\dt$ represent the time step. Note that at a particular point
in time, $v$ and $w$ are functions of $\xi$. We have a Taylor
expansion
\[
f(t+\dt,\xi) = f(t,\xi) + \dt f_t(t,\xi) + \frac{\dt^2}{2}f_{tt}(t,\xi) + \dots
\]
where the task is to replace time-derivatives with derivatives in
space. From \eqref{pde_coeff_gamma} we have
\[
f_t = w(\x,t) - v(\x,t) f_\xi.
\]
This implies that
\begin{align*}
  f_{tt} &= (w(\x,t) - v(\x,t) f_\xi)_t\\
  &= \dd{}{t}w(\x,t) - f_\xi\dd{}{t}v(\x,t)- v(\x,t) f_{t\xi}\\
  &= \dd{}{t}w(\x,t) - f_\xi\dd{}{t}v(\x,t)- v(\x,t) (w(\x,t) - 
  v(\x,t) f_\xi)_\xi\\
  &= \dd{}{t}w(\x,t) - f_\xi\dd{}{t}v(\x,t)- v(\x,t) \left(\dd{}{\xi} w(\x,t) -
    f_\xi \dd{}{\xi}v(\x,t) + v(\x,t) f_{\xi\xi}\right).
\end{align*}
Plugging into the Taylor expansion gives the semi-discrete scheme
\begin{align}
  f(\xi,t+\dt) & = f(\xi,t) + \dt \left(w(\x,t) - v(\x,t) f_\xi\right) + 
  \nonumber\\  
  & + \frac{\dt^2}{2}v(\x,t)\left(-\dd{}{\xi}w(\x,t) + f_\xi\dd{}{\xi}v(\x,t) +
    v(\x,t) f_{\xi\xi}\right) + \nonumber\\
  & + \frac{\dt^2}{2} \left (\dd{}{t}w(\x,t) - f_\xi \dd{}{t}v(\x,t) \right).
  \label{eq:laxw_sd}
\end{align}
Recall that $\x$ corresponds to $f$ via the mapping $\map$.

For the discretization in space we introduce the usual difference
operators $D_0$, $D_-$ and $D_+$, i.e. for some grid function $q_i$ on
the $\xi$-grid, discrete derivative approximations are obtained via
\begin{align*}
  (D_0 q)_i = \frac{1}{2\dxi} (q_{i+1}-q_{i-1}),\quad 
  (D_- q)_i = \frac{1}{\dxi} (q_{i}-q_{i-1}), \quad
  (D_+ q)_i = \frac{1}{\dxi} (q_{i+1}-q_{i}). \quad
\end{align*}
Despite the fact that this Lax-Wendroff method uses centered
difference approximations to discretize a convective equation, it is
stable provided that the CFL condition is satisfied
\cite{LeVeque2002}. We note that we will maintain formal second order
accuracy even if the time derivatives in the last terms are only
approximated to first order and, hence, we use a forward approximation
here by taking an Euler step to get an approximate interface location
where the coefficients can be reevaluated:
\begin{align}
  f^* = f^n + \dt (w(\x^n,t^n) + v(\x^n,t^n)
  D_0{f^n}) \label{eq:laxw_euler_step}
\end{align}
so that
\begin{align}
  f^{n+1} = f^* & + \frac{\dt^2}{2}v(\x^n,t^n) \left(-D_0 w(\x^n,t^n) + 
    (D_0 v(\x^n,t^n))(D_0 f^n)+v(\x^n,t^n) D_+D_- f^n\right) - \nonumber \\
  & + \frac{\dt^2}{2}\left( \frac{w(\x^*,t^{n+1}) - w(\x^n,t^n)}{\dt} -
    \frac{v(\x^*,t^{n+1}) - v(\x^n,t^n)}{\dt}D_0 f^n \right). 
  \label{eq:laxw_method}
\end{align}
We emphasize that the evaluation of the coefficients are typically
done in global coordinates, i.e. via the mapping and its inverse. See
Section \ref{sec:meth_summary} for a summary of the method. As
previously noted, one may often (depending on how $\U$ is computed)
want to apply Strang splitting \eqref{eq:strang}, i.e. taking
\emph{all} time levels above to $t^{n+1/2}$.

\subsection{How to close the system} \label{sec:domain_decomp} 
It remains to discuss the how to couple the segments together. With an
equation of the form \eqref{pde_coeff_gamma} it is clear that
information will be propagated along characteristics across segment
boundaries. In the language of numerical methods for conservation
laws, any difference approximation must then look \emph{upwind} across
to the adjacent segment at any boundary where there are incoming
characteristics.

In our proposed method above, centered differences are used. The
Lax-Wendroff method introduces a diffusive term which stabilizes it
(see LeVeque \cite[pp. 100-102]{LeVeque2002}). Numerically, we then
have information propagating in both directions at each segment
boundary. In practice this means that we need ghost points at each end
of each segment.

To evaluate $D_0$ we can exchange ghost values at time $t^n$. Note
that adjacent segments have different parametrization, so one may need
to construct a small interpolant from the end of the adjacent segment
in order to evaluate the ghost point, see Figure
\ref{fig:ghost_point}. That is, let $(\xi_k,f_k(\xi))$ be the local
representation of segment $k$ (or part of). To view this segment in
the coordinate frame of segment $j$, simply use the mappings:
$(\xi_j,f_j(\xi)) = \map_j \imap_k (\xi_k,f_k(\xi))$. With this
exchange of ghost points, the system is closed in the sense that the
concurrent updates of all segments is equivalent to
\eqref{eq:interface_ode}.

From a theoretical point of view it is important that there exists a
finite overlap between segments (for the manifold properties to
exist). That is, one must be able to go from one segment to another
via the global coordinate representation. If this property is not
satisfied, it becomes impossible to evaluate the ghost points as
discussed above. Thus, it will be necessary as the interface deforms
to expand and/or contract the domain of definition of some, or all,
segments. However, growing a segment involves an adjacent segment in
the same fashion as the ghost points. We emphasize that this is
handled on a regular grid and, thus, does not pose a particular
difficulty (as one may view the point distribution problem in front
tracking methods).

\begin{figure}
  \centering
  \includegraphics{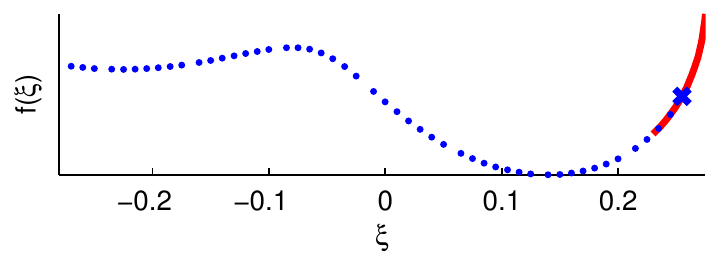}
  \caption{Ghost point $(\times)$, evaluated from part of adjacent
    segment (solid line), with local segment grid function (dots).}
  \label{fig:ghost_point}
\end{figure}

\subsection{Implicit methods for segment advection}
The natural framework for treating hyperbolic PDEs, including the
segment advection equation $f_t + v(\xi,t) f_\xi = w(\xi,t)$, are
explicit finite volume methods (to which the Lax-Wendroff method may
be said to belong). None the less, implicit methods are often proposed
for interface tracking problems, with enhanced time-stability in mind.

As a motivating example, consider a multi-phase flow application. A
common approach is to split the tasks of interface tracking and
solving flow equations into interleaving steps. It is evident that
numerical schemes for both interface tracking and bulk flow should be
governed by CFL-type time-step restrictions, and indeed, a wealth of
numerical methods exist with this property. However, it has been
observed, e.g. in Peskin and Tu \cite{Tu1992}, LeVeque and Lee
\cite{Lee2003}, Le et. al. \cite{Le2006} and Li and Ito
\cite[p. 226]{Li2006}, that this split leads to a stiff system that is
no longer stable in the CFL regime. Thus implicit methods are
advocated to maintain the stability region.

A time-discrete implicit method may take the form
\begin{align*}
  f^{n+1} = \mathcal{F}(f^{n+1},f^n,t,\dots),
\end{align*}
which naturally suggests Newton- or fix-point iteration methods to
solve for $f^{n+1}$. In our framework it is natural to iterate the
segments concurrently, \emph{in lieu} of the coupling between adjacent
segments (previous section). Let the leading super-index of $f^{k,m}$
refer to segment $k$ in the collection that constitutes the interface,
and let the latter index denote time-level or iteration count. Then an
iterative method is outlined as follows: Until convergence,
$\|f^{k,(i+1)} - f^{k,(i)}\|< \epsilon, \forall k$, do
\begin{enumerate}
\item Exchange ghost points between all adjacent segments,
\item For each segment, $k$, compute new iterate, e.g.
  \begin{align*}
    f^{k,(i+1)} \leftarrow \mathcal{G}(f^{k,(i)},f^{k,n},\mathcal{F},\dots),
  \end{align*}
  for some $\mathcal{G}$ that defines the iterative procedure,
\end{enumerate}
and let $f^{k,n+1}$ be the converged iterate for each segment.

\subsubsection{Crank-Nicholson (CN) method for segments} 
\label{sec:cn_meth}

We now proceed to a concrete method along the lines of the previous
section. In \cite{Lee2003} and \cite{Le2006} a Crank-Nicholson method
for \eqref{eq:interface_ode} is suggested,
\begin{align*}
  \x^{n+1} = \x^n + \frac{\dt}{2}\left( \U(\x^n,t^n) +
    \U(\x^{n+1},t^{n+1})\right),
\end{align*}
together with an iterative method for solving this nonlinear
system. This suggests that if we have
\begin{align}
  \dd{f}{t} = Q(t)f(\x,t) + P(t) \label{eq:segment_semi-discrete}
\end{align}
where $Q$ is (possibly) a spatial differential operator, we should
study methods of the form
\begin{align}
  f^{n+1} = \mathcal{F}(f^{n+1},f^n) := f^n + \frac{\dt}{2}\left( Q^n
    f^n + Q^{n+1}f^{n+1} + P^n + P^{n+1}\right). \label{eq:cn_method}
\end{align}
From \eqref{eq:segment_pde_static} we get a semi-discrete equation on
the form \eqref{eq:segment_semi-discrete} with
\begin{align*}
  Q = -v(\x,t) D_0\\
  P = w(\x,t).
\end{align*}
We use the following iterative procedure to solve \eqref{eq:cn_method}
for $f^{n+1}$, which follows the approach presented in \cite{Lee2003}:
\begin{enumerate}
\item Initial guess, $f^{(0)} = f^n - \dt v^n D_0 f^n + \dt w^n$. Set $i = 0$.
\item Evaluate $\U^{(i)}$. This might involve solving flow equations
  (or more).
\item Exchange ghost points between all adjacent segments.
\item Evaluate the residual, 
  \begin{align}
    g(f^{(i)}) = f^{(i)} - f^n - \frac{\dt}{2} ( -v^n(t^n) D_0 f^n 
    -v^{(i)}(t^{n+1}) D_0 f^{(i)} + \label{eq:cn_residual}\\ 
    + w^n(t^n) + w^{(i)}(t^{n+1}) ) \nonumber
  \end{align}
\item If $\| g(f^{(i)})\| <\epsilon$ then terminate and set $f^{n+1}
  = f^{(i)}$. Else increment $i$, compute new $f^{(i)}$ and go to
  step 2. Computing the next iterate should be done cleverly --
  LeVeque\cite{Lee2003} suggests a BFGS or SR1 method.
\end{enumerate}

The BFGS method amounts to a Newton-type iteration where, after
applying the Sherman-Morrison formula, an approximation to the inverse
Jacobian (denoted $B$) is computed concurrently with the solution to
the system (from \cite[Sec. 10.2.5]{Li2006}):

\begin{align*}
  f^{(i+1)} = f^{(i)} - B_n^{(i)} g(f^{(i)})\\
  B_n^{(i+1)} = B_n^{(i)} + \frac{\mu_i s_i s_i^T - s_i y_i^T B_i^n -
    B_n^{(i)} y_i s_i^T}{s_i^Ty_i},
\end{align*}
where 
\begin{align*}
  s_i := -B_n^{(i)} g(f^{(i)}),\quad
  y_i := g(f^{(i+1)}) - g(f^{(i)}),\quad
  \mu_i := 1 + \frac{y_i^T B_n^{(i)}y_i}{s_i^T y_i}.
\end{align*}
Here, as in the explicit method, the coefficients $v^{(i)}$ and
$w^{(i)}$ are evaluated from $f^{(i)}$ in global coordinates via the
mapping \eqref{eq:map} and its inverse. We have found that given the
initial guess computed from the previous time step, this iterative
method is reliably convergent in two to five iterations, which is well
within the practical range, even if $\U$ is expensive to evaluate (see
Section \ref{sec:NS_iim} for more remarks on this).

In practice, we have found the iterative method to relax the CFL time
step restriction present in the explicit method. However, arbitrarily
large time steps can not be taken, in part due to the requirement that
the segments maintain a finite overlap throughout the iterative
procedure (for the evaluation of ghost points). More remarks on the
stability of these methods when applied to the Navier-Stokes equations
are given in Section \ref{sec:NS_iim}.

\subsection{Error analysis}
The truncation error in each of the methods above is $\O(\dxi^2) +
\O(\dt^2) + \O(N_t\dxi^p)$, where $p$ is the order of the
interpolation method used to evaluate the ghost points from adjacent
segments and $N_t := T/\dt$ is the number of time steps taken. One can
show, by constructing a partition of unity, that the accuracy order in
the interface itself is equal to the order obtained for each segment
(i.e. second order).

\subsection{Finding and maintaining segments}
An interesting question naturally arises from the assumptions on our
method: does there exist a decomposition of a curve into segments, and
if so, how do we find it and the mapping parameters in \eqref{eq:map}?

In a continuous setting it is clear from elementary calculus that this
partitioning exists provided sufficient regularity on
$\Gamma$. However, the curve may have, or develop, a interval with
arbitrary curvature. This provides a problem in the discrete setting
because it implies an upper bound on the curvature that can be
reasonably resolved with a given segment discretization. There is also
the question of optimality: partitioning the interface with as few
segments as possible. We do not address the latter issue, because
there is no benefit for our method in having very few segments.

We will also give some remarks on how to verify that the segments, as
the interface deforms, remain valid (i.e. single valued). But fist the
partitioning problem.

\subsubsection{Partitioning the interface} \label{sec:segmentation}

Given a discrete plane curve $(x_i, y_i)$, $i=1\dots N$, the task is
to split the index space into intervals that are monotonous with
respect to some basis vector. To be precise, consider an interval $i =
p, p+1, \dots p+M$ and associate with this segment a mapping
\eqref{eq:map}, i.e. $\map: (x,y) \sra (\xi,f)$. The segment of $\x$
from $p$ to $p+M$ is strictly monotonous with respect to the direction
defined by $\map$ if $\xi_i$, $i = 1,\dots,M$, is strictly monotonous.

A large body of work exists in the field of image processing that aims
to reconstruct geometric features from image data with robust
methods. Some related work to our partitioning problem can be found in
\cite{Pavlidis1974,Ventura1992,Katzir1994,Kolesnikov2007}. However, we
were not able to find a suitable method there or in the references
therein. 

\begin{figure} 
  \centering
  \includegraphics[scale=0.7]{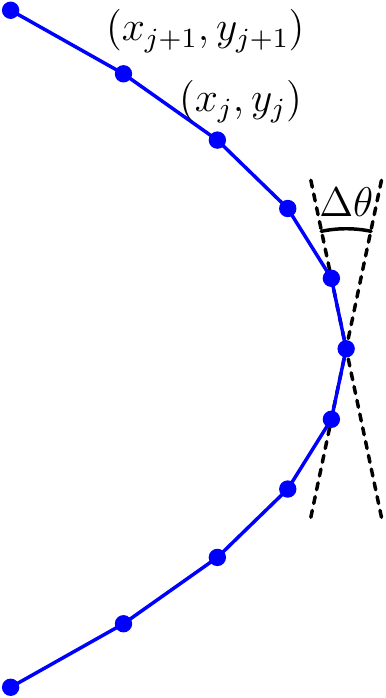}
  \caption{Winding angle, positive with respect to the orientation
    indicated.}
  \label{fig:winding_angle}
\end{figure}

The method we propose uses the \emph{winding angle}, $\wa_i$, as
defined in Figure \ref{fig:winding_angle}, and a cumulative winding
angle,
\begin{align*}
  \theta_I := \sum_{i\in I}^M \wa_i.
\end{align*}
Then we may say that an interval $\varsigma = \{i=p \dots p+M\}$ is
monotonous if
\begin{align}
  \max \theta_\varsigma - \min  \theta_\varsigma < \pi.  
\end{align}
In practice it makes sense to have segments that sweep less than the
maximal winding angle, $\pi$. So we we iterate over the curve, $(x_i,
y_i)$ from some starting index $p$ and break to a new segment when we
encounter an $M$ such that $ \max \theta_\varsigma - \min
\theta_\varsigma \geq \eta$. Here $\eta < \pi$ is a parameter that
affects how many segments we obtain, e.g. $\eta = \pi/2$ gives a
partitioning of a circle into four segments.

\subsubsection{Segment validity} \label{sec:seg_valid} It is clear
that, if the interface deforms sufficiently, the collection of
segments initially constructed may fail to correctly represent the
interface. That is, given an interface, it may not be possible to
satisfy the constraint of having single-valued segment functions with
a given number of segments. Hence, to handle cases with large
deformation, we need to monitor the segment description and recompute
the partition if necessary. We choose to put an upper bound on the
variation of the discrete segment function,
\begin{align*}
  | f_{n+1} - f_n| \leq K |\xi_{n+1} - \xi_n|,
\end{align*}
where $K$ is a parameter. Also, it is sensible to
require that the segments be at least $d$ grid-points long. If the
interface develops a point of very high curvature, the length
condition will suggest that a finer grid is needed to resolve the
interface. Typical values may be $K = 5$ and $d=20$.

\subsection{Summary of method} \label{sec:meth_summary}
Now we summarize the proposed method:
\begin{description}
\item[Time step (explicit method)]
  To advect the interface from $t_n$ to $t_{n+1}$ do,
  \begin{enumerate}
  \item Exchange ghost points between all segments, see Section
    \ref{sec:domain_decomp}.
  \item For each segment (individually): evaluate coefficients
    $v(\x^n,t^n),w(\x^n,t^n)$ by using the transformation to global
    coordinates. Then, take Euler step \eqref{eq:laxw_euler_step} to
    obtain $f^*$, and reevaluate the coefficients
    $v(\x^*,t^{n+1}),w(\x^*,t^{n+1})$. Finally compute the correction
    step \eqref{eq:laxw_method}, getting $f^{n+1}$. See Section
    \ref{sec:laxw}.
  \item Verify that segments are valid, see Section
    \ref{sec:seg_valid}, and possibly grow/contract segments so that
    the evaluation of ghost points possible in the next step.
  \item If segments are not valid, compute a new partitioning
    (Section \ref{sec:segmentation}).
  \end{enumerate}
\end{description}
The implicit method is proceeds similarly, the key difference being
that item two (2.) above is replaced by the iterative procedure given in
Section \ref{sec:cn_meth}.

\subsection{Velocity field decomposition and time-dependent mapping}
\label{sec:vf_decomp}
Finally, before presenting numerical results and applications, we make
an extension of the method, where the mapping parameters are allowed
to vary over time. Is essence, this poses the method hitherto
presented in moving reference frames.

We see in a number of interface tracking applications that there often
exists large scale deformations that translate the interface or rotate
it as a solid body -- physical flows are e.g. often directional with
some length scale. These kinds of motion may be easily treated in our
method by simply updating the segment mapping parameters (the offset
$\b$ and the rotation $\theta$), whereas interface deformations are
treated by solving the advection PDE \eqref{eq:segment_pde}, see
Section \ref{sec:num_res} for numerical results. To make use of this,
consider a decomposition of the velocity field $\U$,
\begin{align}
  \U(\x,t) = \Utrans + \bar{\omega}\times(\x - \b) +
  \Urest(\x,t), \label{eq:vel_decomp}
\end{align}
where $\bar{\omega} = \omega\hat{\mathbf{z}}$. Since we know $\U_i =
(v_i, w_i)$ at many points $\x_i$ we may solve for the decomposition
parameters $\omega$ and $\Utrans = (v_b, w_b)$ in a least squares
sense:
\begin{align}
  \begin{bmatrix}
    (-y_1+y_b) & 1 & 0\\
    (-y_2+y_b) & 1 & 0\\
    \vdots & \vdots & \vdots\\
    (x_1-x_b) & 0 & 1\\
    (x_2-x_b) & 0 & 1\\
    \vdots & \vdots & \vdots
  \end{bmatrix}
  \begin{bmatrix}
    \omega\\
    v_b\\
    w_b
  \end{bmatrix}
  =
  \begin{bmatrix}
    v_1\\
    v_2\\
    \vdots\\
    w_1\\
    w_2\\
    \vdots
  \end{bmatrix}
\end{align}
where $\b = (x_b, y_b)$.

With this we return to \eqref{eq:segment_pde}, calculating
$\dd{}{t}\map\segglob$ where $\dd{\theta}{t} = \omega$ and $\dd{\b}{t}
= \Utrans$ in order to obtain a PDE similar to
\eqref{eq:segment_pde_static}. We get
\begin{align*}
  \dd{}{t}\map(\x) = & \rot(\theta)\dd{\x}{t} + \dd{\b}{t} + 
  \dd{\rot(\theta)}{t}\x\\
  = & \rot(\theta)\U(\x,t) + \Utrans + \dd{\rot(\theta)}{t} \x
\end{align*}
Differentiating the rotation operator is straight-forward,
\begin{align*}
  \dd{}{t}\rot(\theta) =& \begin{bmatrix}  -\dd{\theta}{t}\sin(\theta) & 
    -\dd{\theta}{t}\cos(\theta)  \\ \dd{\theta}{t}\cos(\theta) & 
    -\dd{\theta}{t}\sin(\theta) \end{bmatrix}\\
  =&\,\, \omega \begin{bmatrix} \cos(\pi/2+\theta) & -\sin(\pi/2+\theta) \\ 
    \sin(\pi/2+\theta) & \cos(\pi/2+\theta)\end{bmatrix}\\
  =&\,\, \omega \rot(\pi/2+\theta).
\end{align*}
Using this we get
\begin{align}
  \dd{}{t}\map(\x) = \rot(\theta) \U(\x,t) + \Utrans +
  \omega\rot(\pi/2+\theta)\x.
\end{align}
Inserting this into \eqref{eq:segment_pde} gives a PDE, of the same
type as \eqref{eq:segment_pde_static}, for the segment dynamics in the
case when the velocity field decomposition \eqref{eq:vel_decomp}:
\begin{align}
  \pd{f}{t} + \big( \rot(\theta) \Urest + \Utrans +
  \omega\rot(\pi/2+\theta)\segglob \big) \cdot \left[\pd{f}{\xi}\quad
    -1\right] = 0. \label{eq:segment_pde_dynamic}
\end{align}
The discretization of this PDE fits naturally into the explicit method
summarized above, with two additions. First, apply Strang splitting
\eqref{eq:strang} to obtain a second order accurate method. Secondly,
the mapping parameters are updated before the coefficients $(v,w)$ are
reevaluated,
\begin{align*}
  \theta^{n+1} = \theta^n + \dt \omega\\
  \b^{n+1} = \b^n + \dt \Utrans.
\end{align*}

\section{Numerical results for interface tracking} 
\label{sec:num_res}
Here we provide numerical results for the pure interface tracking
problem. So as to not introduce any new discretization errors, we let
the convective velocity field be given by a closed expression (as
opposed to computing an approximate solution to e.g. Navier Stokes
equations):
\begin{align}
  \U_I(\x, t) = \cos(\pi t)\big( -\cos(x_i- \pi/2) \sin(x_2 - \pi/2)\quad
  \cos(x_2 - \pi/2) \sin(x_1 - \pi/2)\big)\label{eq:vf_osc}\\
  \U_{II}(\x, t) = \big( \sin(\pi x)^2\sin(2\pi y)\quad
  -\sin(\pi y)^2\sin(2\pi x)\big)\label{eq:vf_vortex_II}
\end{align}
An example of a computation with $\U = \U_{II}$ can be seen in Figure
\ref{fig:run1}. The curvature of the interface is known to to be
difficult to compute, even to ``eye norm'' accuracy. It is even the
case with many methods that some kind of filtering is needed to get
acceptable curvature results, though this is not generally
discussed. In Figure \ref{fig:run1}, we also show the interface
curvature at different times. The two sharp peaks are well captured,
without any oscillations, and there are no visible artifacts at the
segment boundaries. We compute curvature in local coordinates, via
\begin{align}
  \kappa(\xi) = \frac{f''(\xi)}{(1+ f'(\xi)^2)^{3/2}}, 
  \label{eq:curvature_local} 
\end{align} 
as discretized with the usual centered difference approximations.
\begin{figure}
  \begin{center}
    \includegraphics[width=.28\columnwidth]{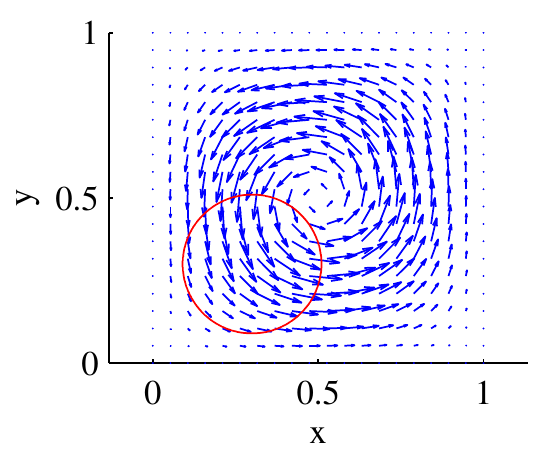}
    \includegraphics[width=.28\columnwidth]{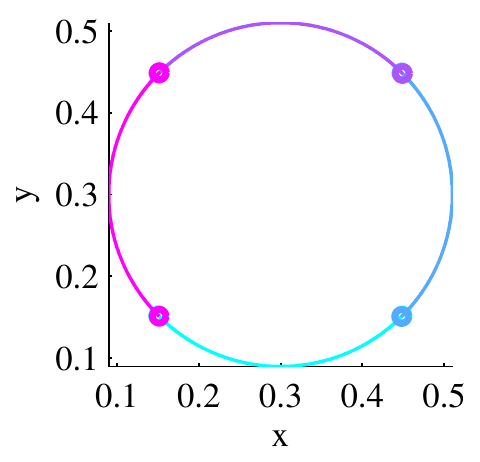}
    \includegraphics[width=.28\columnwidth]{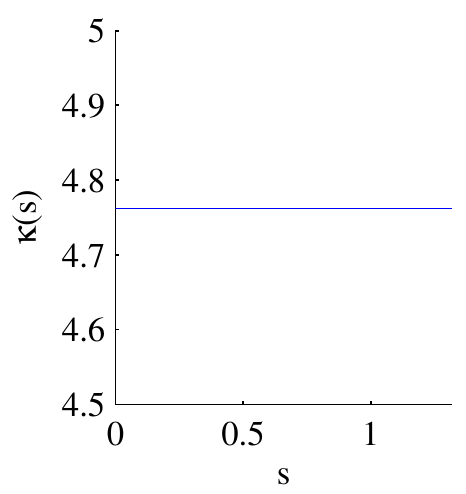}
    \includegraphics[width=.28\columnwidth]{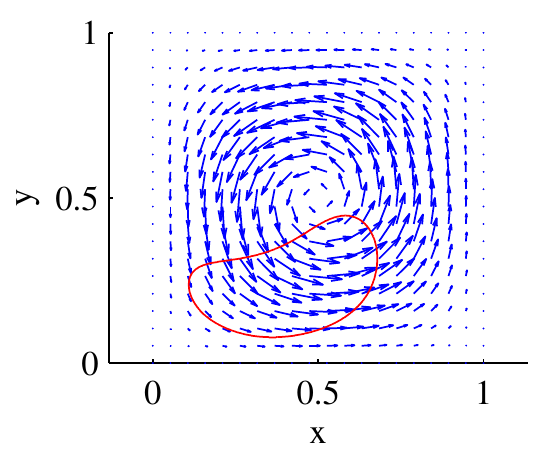}
    \includegraphics[width=.28\columnwidth]{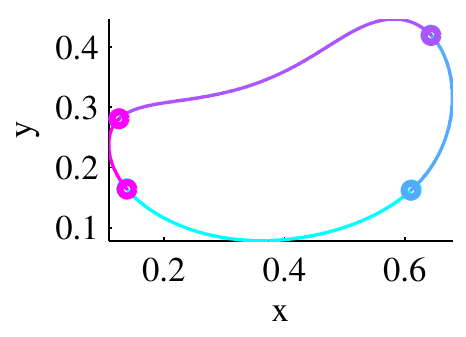}
    \includegraphics[width=.28\columnwidth]{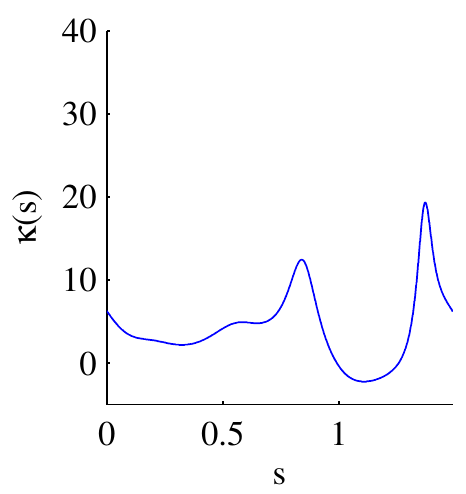}
    \includegraphics[width=.28\columnwidth]{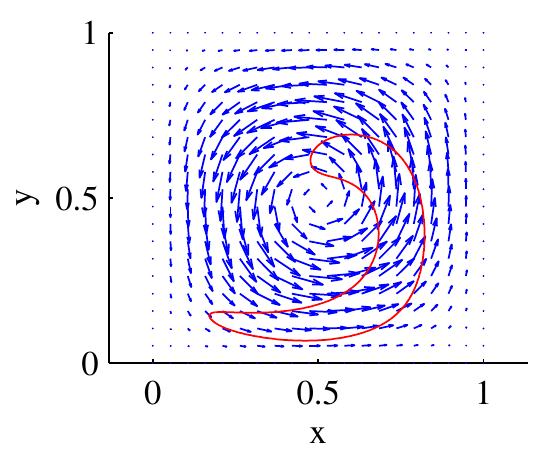}
    \includegraphics[width=.28\columnwidth]{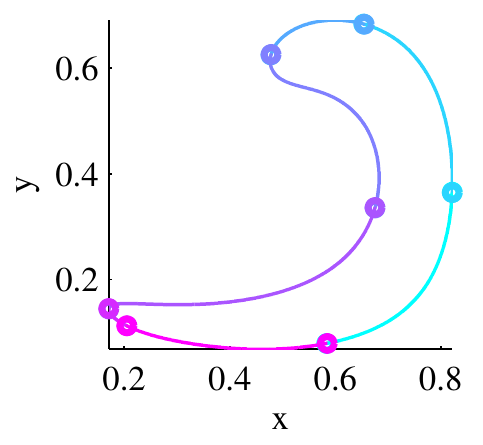}
    \includegraphics[width=.28\columnwidth]{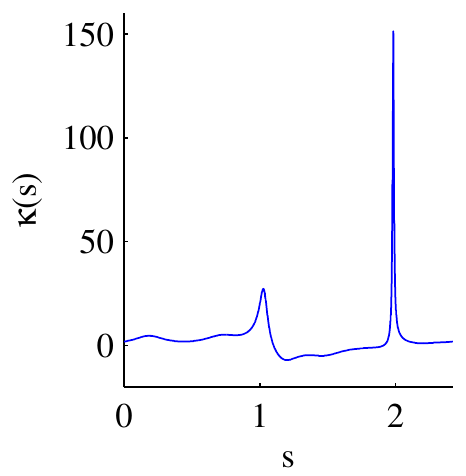}
    \includegraphics[width=.28\columnwidth]{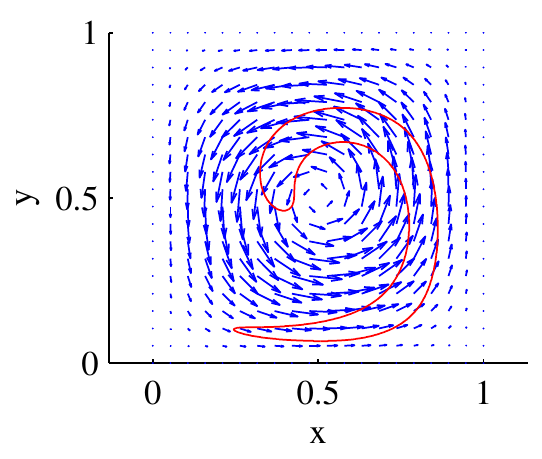}
    \includegraphics[width=.28\columnwidth]{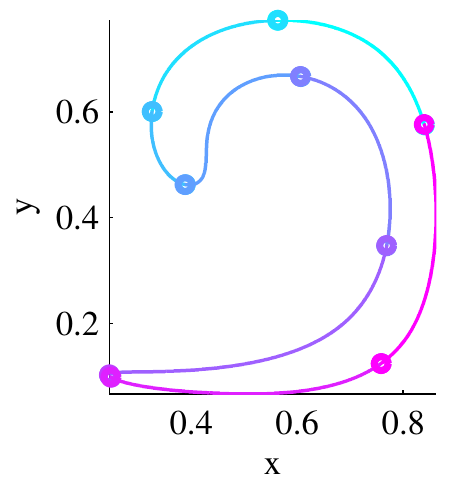}
    \includegraphics[width=.28\columnwidth]{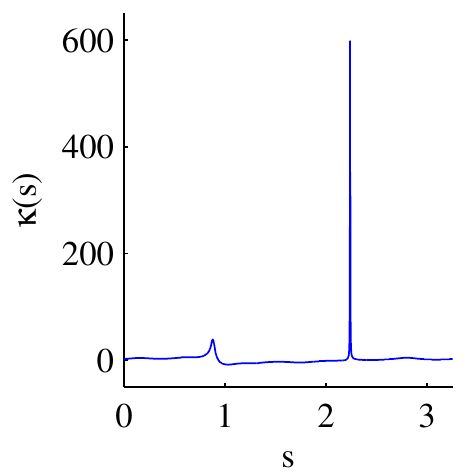}
  \end{center}
  \caption{Interface and curvature at times $t = 0,\, 0.2,\, 0.6,\,
    0.9$, where the convective velocity field is given by
    \eqref{eq:vf_vortex_II}. The circles mark (middle column) the
    segment boundaries.} 
  \label{fig:run1}
\end{figure}

For testing numerical convergence we construct test cases that have an
analytical solution. The convective field $\U_I$ oscillates so that,
at $t = 1$, the interface is returned to its initial configuration,
which we take to be a circle. In Figure \ref{fig:conv1} we present
convergence results for both the interface location and its curvature
as the grid is refined, with the explicit (LaxW) method. The time step
is refined so that the ratio $\dxi/\dt=2$ is kept constant. We see the
expected second order convergence in both time and space for the
interface, in $\|\x(T) - \x(0)\|_\infty$. In Figure \ref{fig:conv2}
we give similar results for the implicit (CN) method.

\subsection{Convergence in curvature}
Also in Figure \ref{fig:conv1}, to substantiate our claim that the
interface curvature is well captured by our method, we give
convergence results for curvature. Here we see first order convergence
in $\| \kappa(s,T) - 1/r \|_\infty$ and close to second order
convergence in $\| \kappa(s,T) - 1/r \|_{l^2}$. That the 2-norm
convergence is substantially higher than in $\infty$-norm tells us
that the first order errors are highly localized. These results are
with the LaxW method and we wish to emphasize that no filtering or
smoothing of any kind were used to obtain them. Together with the
second order convergence of the interface in $\infty$-norm, we take
these curvature convergence results as strong endorsement of the
accuracy of the proposed method. In particular, the
domain-decomposition approach seems vindicated.

With the CN method we were not able to get equally convincing
convergence results for the curvature. Still, in Figure
\ref{fig:conv2}, we have first order convergence of the curvature in
2-norm (in $\infty$-norm we have convergence, but of order $<1$).

\begin{figure}
  \centering
    \includegraphics[width=.28\columnwidth]{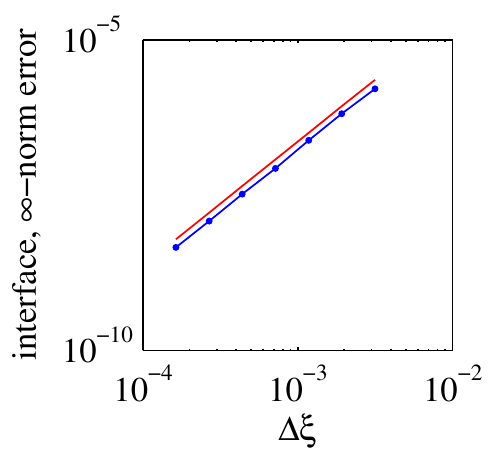}
    \includegraphics[width=.28\columnwidth]{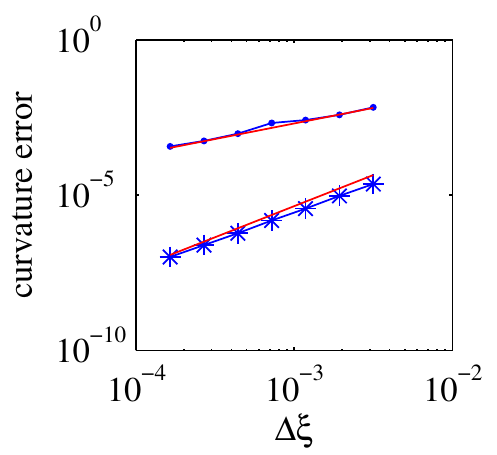}
    \includegraphics[width=.28\columnwidth]{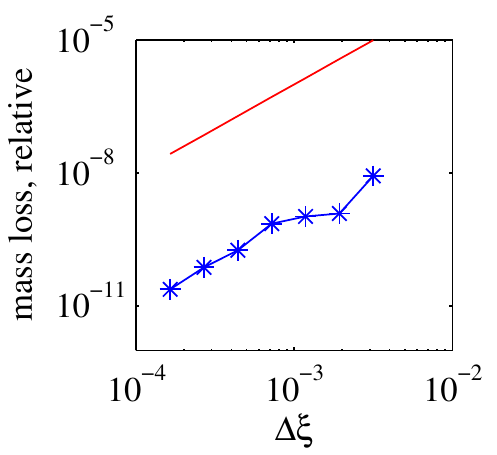}
  \caption{Explicit (LaxW) method. Left: Convergence of interface in
    $\infty$-norm as grid is refined, with $\U = \U_I$
    (Eq. \eqref{eq:vf_osc}), at $t=1$. Middle: Convergence of
    interface curvature in ($\cdot$) $\infty$- and ($*$)
    $2$-norm. Right: Convergence in relative mass loss (solid line
    denotes second order convergence).}
  \label{fig:conv1}
\end{figure}

\begin{figure}
  \centering
    \includegraphics[width=.28\columnwidth]{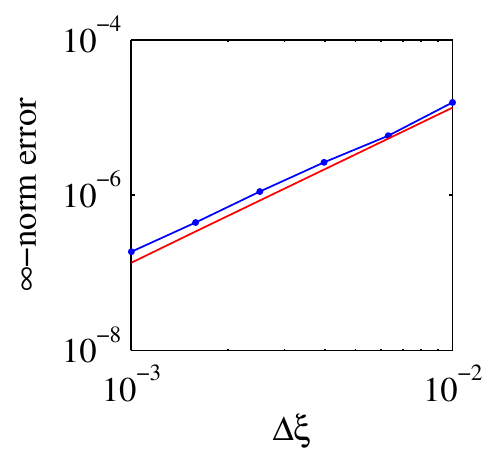}
    \includegraphics[width=.28\columnwidth]{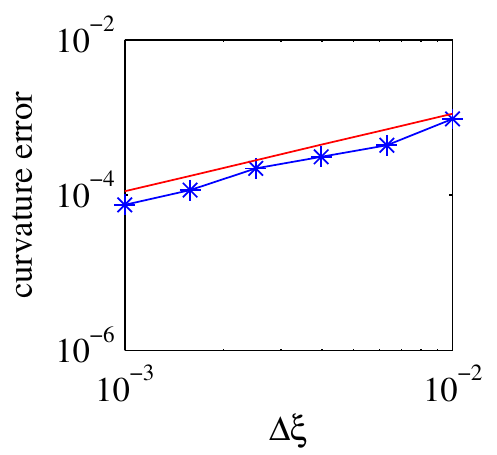}
    \includegraphics[width=.28\columnwidth]{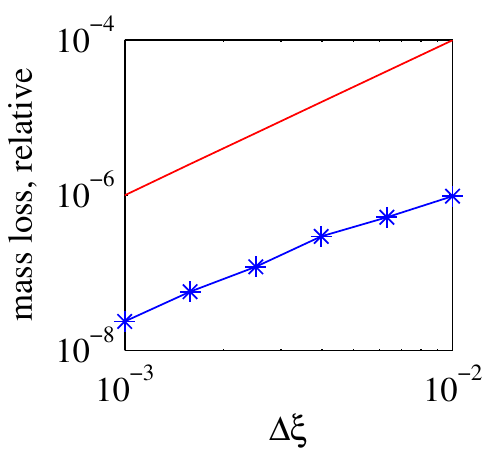}
  \caption{Implicit (CN) method. Left: Convergence of interface in
    $\infty$-norm as grid is refined, with $\U = \U_I$. Middle:
    Convergence of interface curvature in ($*$) $2$-norm. Right:
    Convergence in relative mass loss (solid line denotes second order
    convergence). }
  \label{fig:conv2}
\end{figure}

\subsection{Mass, or area, conservation}
There is no inherent mass conservation in our method. On the other
hand, there is no reason to expect it to suffer from poor mass
conservation in the same way as level set methods are known to do. 

In Figures \ref{fig:conv1} and \ref{fig:conv2} (right column) we give
mass loss results as the interface is refined, for the explicit and
implicit method. That is, we compute the mass loss as a relative
measure, via $(\text{mass}(0)-\text{mass}(T))/\text{mass}(0)$ where
the mass is simply computed as the area of the polygon enclosed by the
interface.

We note here that the mass loss shows second order convergence as the
interface is refined, and that the constant is very small -- several
orders of magnitude below unity.

\subsection{Dynamic mapping parameters}
Finally, before we develop applications, we give numerical results for
the extended method in Section \ref{sec:vf_decomp}, where the mapping
parameters vary. Again with the oscillating circle case $\U = \U_I$,
we have second order convergence in the interface in $\infty$-norm,
first order convergence in curvature and second order convergence in
the mass conservation, see Figure \ref{fig:conv3}. In Figure
\ref{fig:rot_1} we give results where the interface is advected in a
pure rotation velocity field, as is a common test case for interface
tracking methods. A notable characteristic of our method is that,
aside from initialization errors, there are no numerical errors in
this computation, because of the velocity field decomposition.

\begin{figure}
  \centering
  \includegraphics[width=.28\columnwidth]{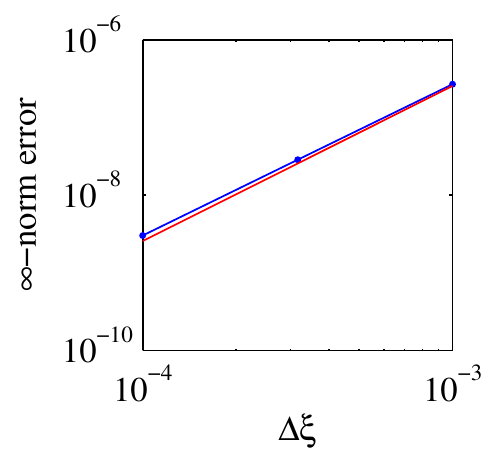}
  \includegraphics[width=.28\columnwidth]{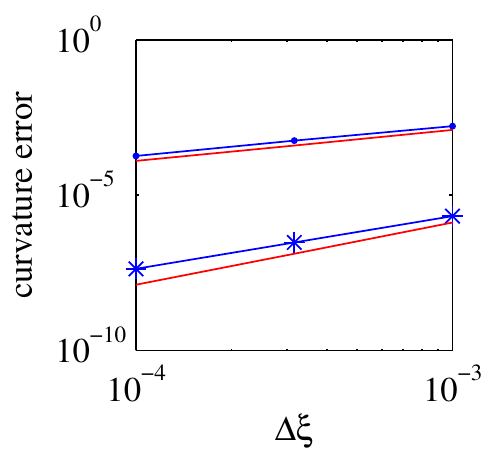}
  \includegraphics[width=.28\columnwidth]{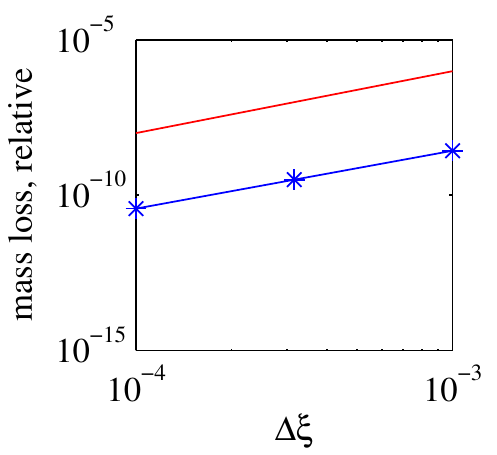}

  \caption{Explicit (LaxW) method, with velocity field decomposition
    (Section \ref{sec:vf_decomp}). Left: Convergence of interface in
    $\infty$-norm as grid is refined, with $\U = \U_I$. Middle:
    Convergence of interface curvature in ($\cdot$) $\infty$- and
    ($*$) $2$-norm. Right: Convergence in relative mass loss (solid
    line denotes second order convergence).}
  \label{fig:conv3}
\end{figure}

\begin{figure}
  \centering
  \includegraphics{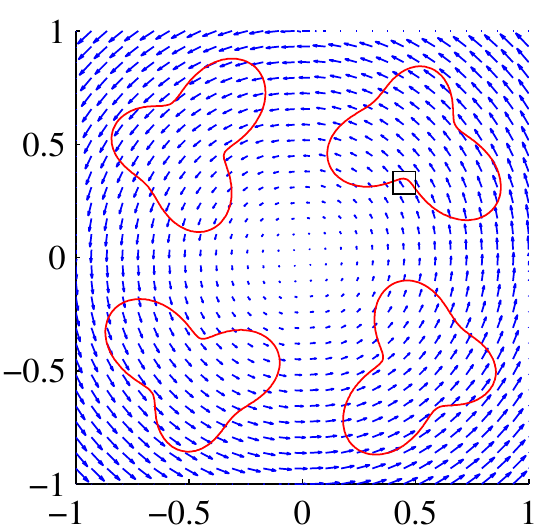} 
  \includegraphics{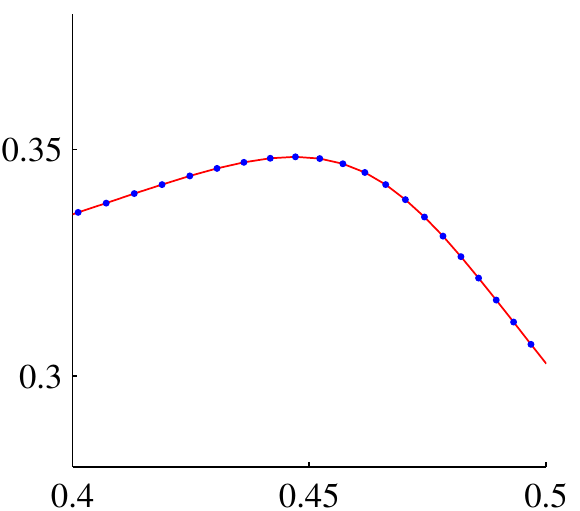}
  \caption{Left: Interface at four times, and pure rotation velocity
    field. Left: Close up of boxed region in left figure, with
    interface at $t=0$ as solid red line, and interface at $t=2\pi$ as
    dots. These are indistinguishable.}
  \label{fig:rot_1}
\end{figure}

\input{bie} 
\input{ns}

\section{Conclusions and outlook}

We have introduced an interface tracking method which is both Eulerian
and explicit, as motivated by current and future needs for micro- and
complex flow modeling and simulation. For instance, the PDE-based
methods by Khatri and Tornberg \cite{Khatri2009} for simulating
soluble and insoluble surfactants fit naturally into this framework.

The method presented is based on a domain decomposition approach,
where the interface is split into segments, or patches. On each patch
a hyperbolic PDE is solved in time to update the interface location,
and adjacent segments are coupled via an exchange of ghost points. One
explicit and one implicit time step method was presented, both second
order accurate.

Numerically, the accuracy and consistency order of the proposed method
was established: second order in $\infty$-norm for the interface
location. We were also able to present first and second order
convergence in the interface curvature -- a quantity which is known to
be hard to compute (without filtering). Mass conservation was found to
be well behaved, converging at second order with a constant several
orders of magnitude below unity.

An extension of the method was also presented, which made use of a
decomposition of the velocity field into translation and rotation
components, as motivated by a variety of flow applications where such
components tend to dominate. The resulting method was also shown
numerically to have the desirable accuracy characteristics mentioned
above.

We treated two applications: Stokes and Navier-Stokes flow. The first
used a boundary integral method, and illustrated how an application
with integration over the interface can be treated in our
framework. We found that it was simple to derive and analyze a method
which was second order accurate for integrals of the Greens' function
for Stokes flow in free space.

Finally, with the application of our method to two-phase
incompressible Navier Stokes flow, we showed that our method works
well with existing components for such problems. Both the immersed
interface- and immersed boundary method was used. We gave and
evaluated a first order accurate method for solving the Navier-Stokes
equations and presented computations of a bubble in shear flow. The
two-phase flow results are by no means new, but serve to underline
that our method can be used in established applications with good
results.

However, as of yet, we have not generalized our method to 3D in
sufficient generality. The framework and equations generalize
naturally -- the difficulty is representing closed surfaces as a union
of patches, in a way that allows the patches to couple together
robustly. One would like to use regular grids in 2D for these patches,
to be able to use well-established numerical method for advecting the
interface, computing curvature and quadrature etc.

None the less, we believe that at a method that is explicit in the
interface location and Eulerian is very useful for treating complex
interface tracking applications. There is a wealth of interesting
flow applications on the micro scale where continuum modeling on the
interface becomes relevant.

\section*{References}
\bibliographystyle{elsarticle-num}
\bibliography{pug}{}

\end{document}

%% file: defs.tex
\usepackage{amsmath, amssymb}

\newcommand{\R}{\mathbb{R}}

\newcommand{\pd}[2]{\frac{\partial #1}{\partial #2}}
\renewcommand{\d}{\mathrm{d}}
\newcommand{\dd}[2]{\frac{\mathrm{d} #1}{\mathrm{d} #2}}

\newcommand{\sra}{\rightarrow}

\newcommand{\x}{\mathbf{x}}
\renewcommand{\b}{\mathbf{b}}
\newcommand{\U}{\mathbf{u}}
\newcommand{\F}{\mathbf{F}}
\newcommand{\f}{\mathbf{f}}
\newcommand{\Utrans}{\mathbf{u}_\text{trans}}
\newcommand{\Urest}{\mathbf{u}_\text{rest}}
\newcommand{\nhat}{\hat{n}}

\newcommand{\dt}{\Delta t}
\newcommand{\dxi}{\Delta \xi}
\newcommand{\wa}{\Delta \theta}

\newcommand{\map}{\mathcal{M}}
\newcommand{\imap}{\mathcal{L}}
\newcommand{\rot}{\mathcal{T}}
\newcommand{\jac}{\mathcal{J}}

\newcommand{\iface}{\Gamma}
\newcommand{\segglob}{\gamma}
\newcommand{\segloc}{\gamma_\text{loc}}

\newcommand{\RE}{\text{Re}}
\newcommand{\CA}{\text{Ca}}
\newcommand{\jmp}[1]{[#1]}
\renewcommand{\O}{\mathcal{O}}


%% file: bie.tex
\section{Boundary integral method for Stokes in 2D} 
\label{sec:BIE_Stokes}

\newcommand{\hx}{\hat{x}}
\newcommand{\hn}{\hat{n}}

When applicable, boundary integral methods are often efficient. Here
we give a boundary integral formulation for Stokes flow in 2D using
our method. This illustrates some important simplifications that arise
when using our method.

For an introduction to the field of boundary integral methods for
viscous flow, we refer to the textbooks Pozrikids
\cite{Pozrikidis1992} and Kim and Karrila \cite{Kim1991}, as well as a
more recent survey paper by Pozrikidis \cite{Pozrikidis2001}. 

Using the so called single layer formulation, the velocity at any
point, $x_0\in\Omega \subset \R^n$, can be evaluated as
\begin{align}
  u_i(x_0) = \int_\Gamma G_{ij}(x,x_0) q_j(x) \d S(x), \label{eq:BIE_u}
\end{align}
where $G_{ij}$ is a Greens function (fundamental solution) called the
Stokeslet, and $q_j$ is a source density on the interface. Here we
consider a free space problem in $\R^2$. Then the Stokeslet is given by
\begin{align*}
  G_{ij}(x,x_0) = -\delta_{ij} \log(r) + \frac{\hx_i \hx_j}{r^2},
\end{align*}
where $\hx = x - x_0$, $r = |\hx|$ and $\delta_{ij}$ is the Kronecker
symbol. There remains only to compute the densities $q_j$, which in
general requires solving a (full) linear system. However, in the case
where the viscosity inside is equal to the viscosity outside of the
interface, one obtains
\begin{align*}
  q_j(s) = -\frac{1}{4\pi\mu} f_j(s),
\end{align*}
where $s$ parametrizes $\Gamma$ and $\mu$ is the viscosity. These
results were given in \cite{Pozrikidis1990}. The force acting on the
interface is simply
\begin{align*}
  f_j(s) = \sigma \kappa(s) \hn_j(s),
\end{align*} 
where $\sigma$ is the coefficient of surface tension, $\kappa$ is the
curvature and $\hn_j$ are the unit normals. In terms on
non-dimensional variables, we write
\begin{align*}
  q_j(s) =-\frac{1}{4\pi\CA} \kappa(s) \hn_j(s),  
\end{align*}
where $\CA:=\frac{U\mu}{\sigma}$ is known as the capillary number (for
some typical velocity scale $U$).

Now we pose the evaluation of \eqref{eq:BIE_u} in the setting of our
segments. By definition we have that any integral over the interface
can be split into a sum of integrals over the segments $\gamma_i$,
\begin{align} 
  \int_\Gamma g(s) \d s & = \sum_i \int_{\gamma_i} \tilde{g}(\xi) \d\xi_f 
  \label{eq:bie_genral_integral}\\
  & = \sum_i \int_{\gamma_i} g(s(\xi)) \sqrt{1+f'(\xi)^2} \d \xi \nonumber.
\end{align}
In order to perform this change of variables in \eqref{eq:BIE_u} from
a global parametrization, $s$, to local variables $(\xi, f(\xi))$ we
note a few invariant properties of the integrand. First, the curvature
is a geometric invariant, so instead of computing $\kappa(s)$ we can
compute it in local coordinates directly, as given by
\eqref{eq:curvature_local}. Similarly, for the normal vectors we
directly compute
\begin{align*}
  \hn(\xi) = [f'(\xi) \quad -1].
\end{align*}
Finally, note that the global-to-local mapping, $\map$
(cf. Eq. \eqref{eq:map}) is distance preserving. Hence, we directly
express the Stokeslet in local coordinates
\begin{align*}
  G_{ij}(\xi, f(\xi), x_0) = -\delta_{ij} \log(r) + \frac{\tilde{\x}_i
    \tilde{\x}_j}{r^2},
\end{align*}
where $\tilde{\x} := (\xi, f(\xi)) - \map x_0$ and $r=\|
\tilde{\x}\|$. For a point $x_0$ away from the interface, using the
inverse mapping \eqref{eq:imap}, we have that
\begin{align}
  u(x_0) & = -\frac{1}{4\pi \mu} \sum_k \imap_k \int_{\gamma_k}
  \kappa(\xi)
  G_{ij}(\xi, f(\xi), x_0) \hn_j(\xi) \d \xi_f \nonumber \\
  & = -\frac{1}{4\pi \mu} \sum_k \imap_k \int_{\gamma_k}
  \frac{f''(\xi) }{1+f'(\xi)^2} G_{ij}(\xi, f(\xi), x_0) \hn_j(\xi)\d
  \xi \label{eq:BIE_u_seg}\\
  & = -\frac{1}{4\pi \mu} \sum_k \imap_k \int_{\gamma_k} q^f_j(\xi)
  G_{ij}(\xi, f(\xi), x_0)\d \xi. \nonumber
\end{align}
These integrals are naturally discretized on the segment grids, where
the start of the integration interval, denoted $\xi'$, corresponds to
the end of the previous segment, i.e we compute integrals of the form
\begin{align*}
 \int_{\gamma_k} g(\xi) d\xi = \int_{\xi'}^b g(\xi) d\xi.
\end{align*}
These are discretized with trapezoidal quadrature over the $\xi$-grid.

\subsection{Treatment of singular integrands}
As the point $x_0$ approaches the interface the integrands in
\eqref{eq:BIE_u_seg} become singular. In order to evaluate $u_j$ on
the interface (e.g. for time-dependent problems) we need to provide
precisely how to compute these integrals -- showing that they exist
and what the relevant limits are.

Suppose $x_0$ lies on the interface. On some segment, $x_0$
corresponds to a point in local coordinates $(\xi_0,f(\xi_0))$. The
distance function then takes the form
\begin{align}
  r = r(\xi,\xi_0,f) = \sqrt{ (\xi - \xi_0)^2 + (f(\xi) -
    f(\xi_0))^2}.
\end{align}

First, we have the $\log$-type singularity due to the first term in
$G$, which needs careful treatment. Secondly, the terms involving $x_i
x_j/r^2$ are non-singular, but the limits as $\xi\rightarrow \xi_0$
need to be determined.

The logarithmic singularity can be subtracted away, in two steps:
\begin{align}
  u_j =& \int_a^b q_j(\xi) \log(r(\xi,\xi_0,f)) d \xi \nonumber \\
  & = \int_a^b [q_j(\xi) -q_j(\xi_0)] \log(r(\xi,\xi_0,f)) d \xi +
  q_j(\xi_0) \int_a^b \log(r(\xi,\xi_0,f)) d \xi \nonumber\\
  =& \int_a^b [q_j(\xi) -q_j(\xi_0)] \log(r(\xi,\xi_0,f)) d \xi +\nonumber \\
  &+q_j(\xi_0) \left[ \int_a^b \log\left( \frac{
        r(\xi,\xi_0,f)}{|\xi-\xi_0|} \right) d \xi + \int_a^b
    \log(|\xi-\xi_0|) d\xi \right].\label{eq:bie_singularity}
\end{align}
Now we show that the first two integrands are non-singular, compute
the corresponding limits as $\xi \rightarrow \xi_0$ and compute the
last integral analytically. For the latter we get
\begin{align}
  \int_a^b \log(&|\xi-\xi_0|) d\xi = \\
  =&
  \begin{cases}
    (\log(|b-x_0|) -1)(b-x_0) & \text{if } a = x_0\\
    -(\log(|a-x_0|) -1)(a-x_0) & \text{if } b = x_0\\
    (\log(|b-x_0|) -1)(b-x_0)-(\log(|a-x_0|) -1)(a-x_0)& \text{otherwise}. 
    \nonumber
  \end{cases}
\end{align}
The integrand in the first integral in \eqref{eq:bie_singularity}
behaves as $x\log x$ for small $x$, and hence goes to zero. More
specifically, the limits of the two integrands (i.e. for the x- and
y-components of $u$) are
\begin{align*}
  \lim_{\xi\rightarrow \xi_0} (q^f_1(\xi) - q^f_1(\xi_0)) \log(r(\xi)) = 
  \lim_{\xi\rightarrow \xi_0} \left(
    \frac{f'(\xi) f''(\xi)}{ 1+f'(\xi)^2} - 
    \frac{f'(\xi_0) f''(\xi_0)}{ 1+f'(\xi_0)^2} 
  \right) \log r(\xi) = 0
\end{align*}
and
\begin{align*}
  \lim_{\xi\rightarrow \xi_0} (q^f_2(\xi) - q^f_2(\xi_0)) \log(r(\xi)) = 
  \lim_{\xi\rightarrow \xi_0} \left(
    \frac{- f''(\xi)}{ 1+f'(\xi)^2} - 
    \frac{- f''(\xi_0)}{ 1+f'(\xi_0)^2} 
  \right) \log r(\xi) = 0.
\end{align*}
For the second integral in \eqref{eq:bie_singularity} the relevant
limit is
\begin{align*}
  \lim_{\xi\rightarrow\xi_0} \frac{
    r(\xi,\xi_0,f)}{|\xi-\xi_0|} = \sqrt{1+f'^2(\xi_0)} > 0,
\end{align*}
so that the second integrand goes to $\log(\sqrt{1+f'^2(\xi_0)})$ near
$\xi = \xi_0$. This concludes the treatment of the integrals involving
$\log$-singular integrands. The simple quadrature rule above is
applied and the limit values are used when the integrands are not
immediately well defined.

Similarly, for the terms in $G$ involving $\x_i \x_j/r^2$ we provide
the required limits,
\begin{align*}
  \lim_{\xi \rightarrow \xi_0} \frac{(\xi - \xi_0)^2}{r(\xi)^2} =
  \frac{1}{1 + f'(\xi_0)^2}\\
  \lim_{\xi \rightarrow \xi_0} \frac{(\xi - \xi_0)(f(\xi) -
    f(\xi_0))}{r(\xi)^2} =
  \frac{f'(\xi_0)}{1 + f'(\xi_0)^2}\\
  \lim_{\xi \rightarrow \xi_0} \frac{(f(\xi) -
    f(\xi_0))^2}{r(\xi)^2} =
  \frac{f'(\xi_0)^2}{1 + f'(\xi_0)^2}.
\end{align*}
This concludes the formulation of the boundary integral method.

\subsection{Numerical results for BIE Stokes}
The boundary integral method above fits into the proposed interface
tracking method, simply by letting \eqref{eq:BIE_u_seg} define the
convective field, $\U$, in e.g. the explicit (LaxW) time step scheme.

\begin{figure}
  \centering 
  \includegraphics[width=.4\columnwidth]{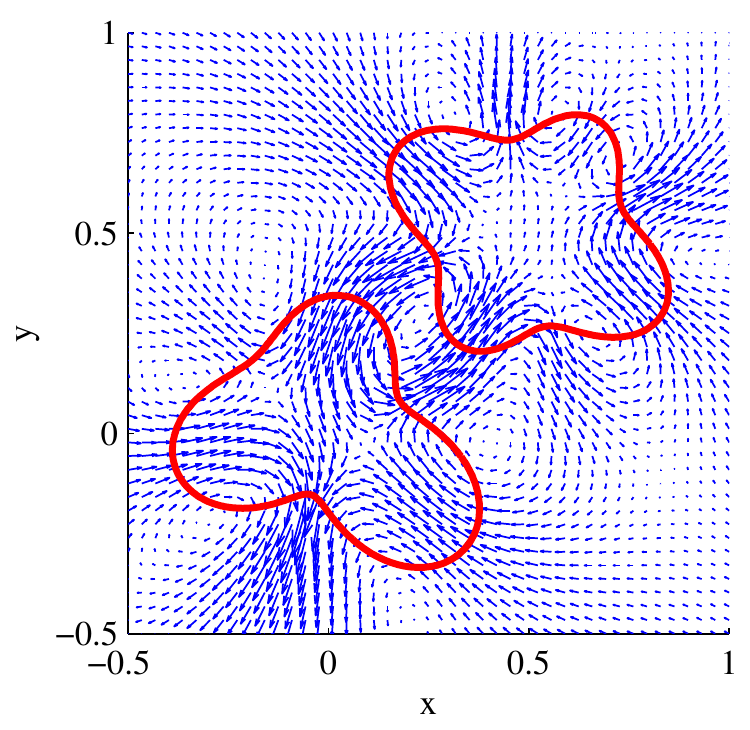}
  \includegraphics[width=.4\columnwidth]{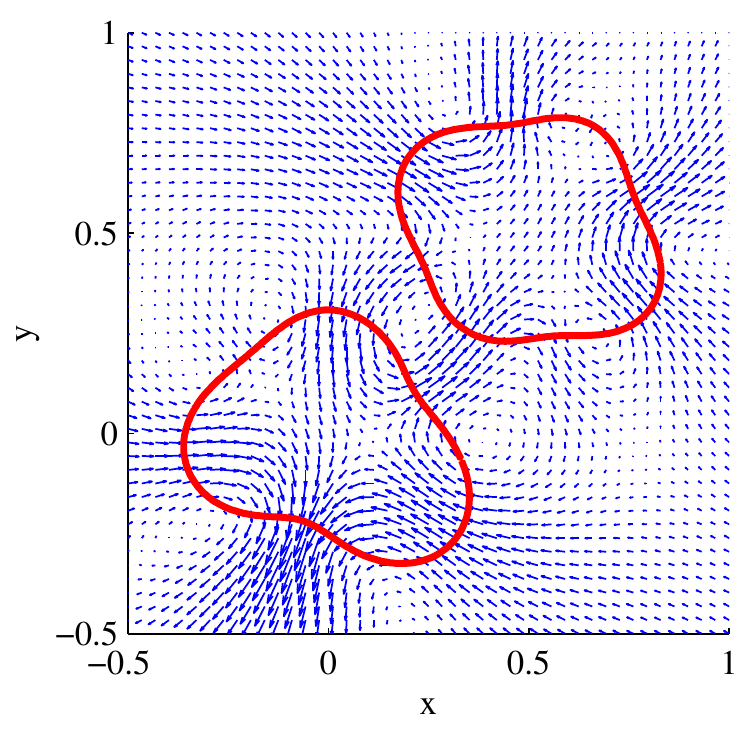}
  \includegraphics[width=.4\columnwidth]{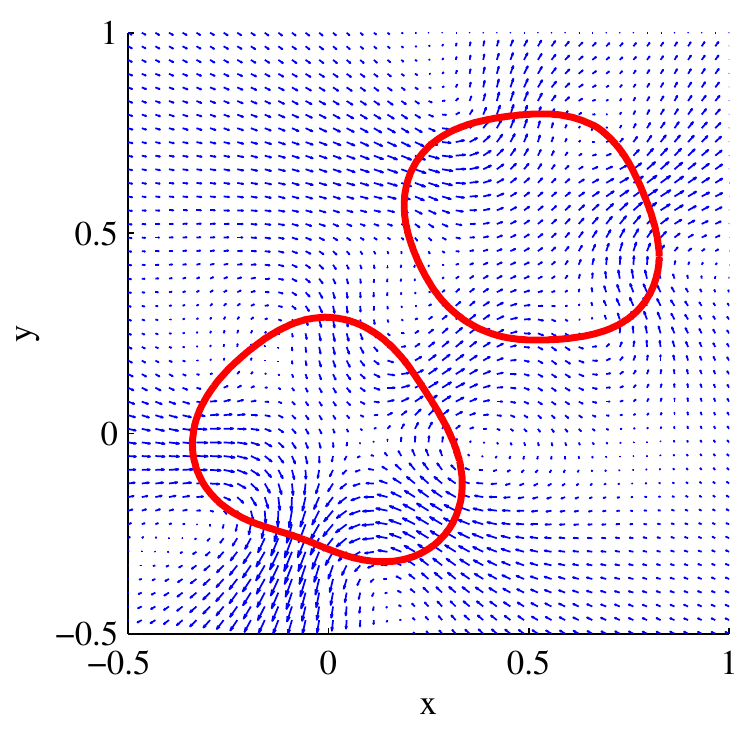}
  \includegraphics[width=.4\columnwidth]{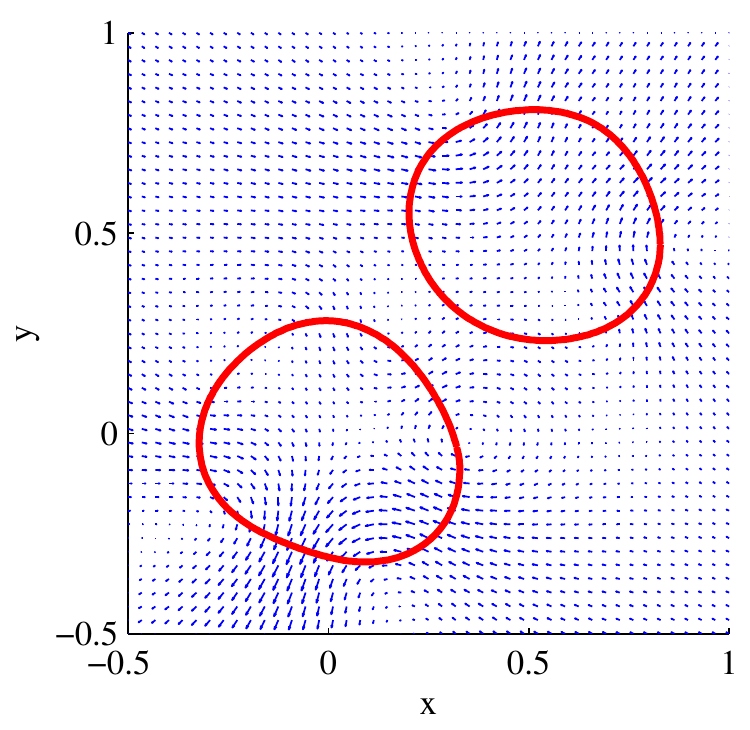}
  \caption{Interface in free space Stokes flow, driven by surface
    tension, at four times. The convective field $\U$ is obtained by
    evaluation the (singular) integrals in \eqref{eq:BIE_u_seg} on the
    interface. The flow fields shown are evaluations of
    \eqref{eq:BIE_u_seg} on a grid, used for visualization only.}
  \label{fig:bie_001}
\end{figure}

In Figure \ref{fig:bie_001} we give illustrative plots of the flow
field, evaluated over a grid, as computed via the boundary integral
\eqref{eq:BIE_u_seg}. These show the expected flow features for a free
space flow driven by the curvature of the interface. For a comparison
to Navier-Stokes flow, and results for Stokes for a shear flow, see
Section \ref{sec:comp} and Figure \ref{fig:ns_shear_1}. Note that
evaluating the velocity field away from the interface is only done for
post processing, e.g. visualization. The resolution here was
$\dxi=1/100$, and the interface was initialized with four segments (as
in Figure \ref{fig:interface_mark_seg}).

\begin{figure}
  \centering
  \includegraphics[width=.4\columnwidth]{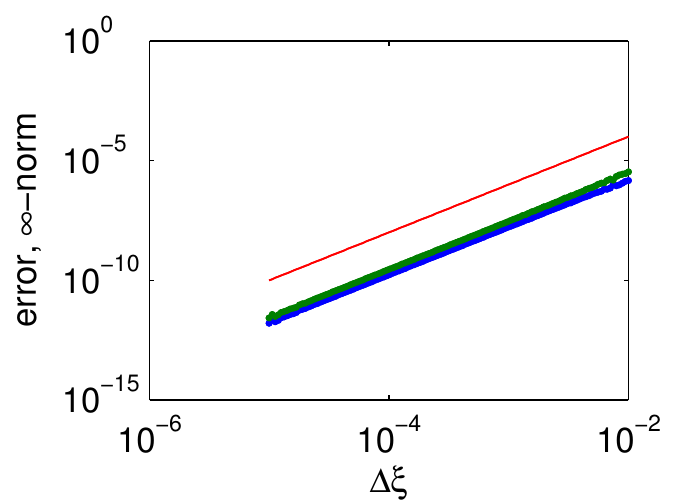}
  \includegraphics[width=.4\columnwidth]{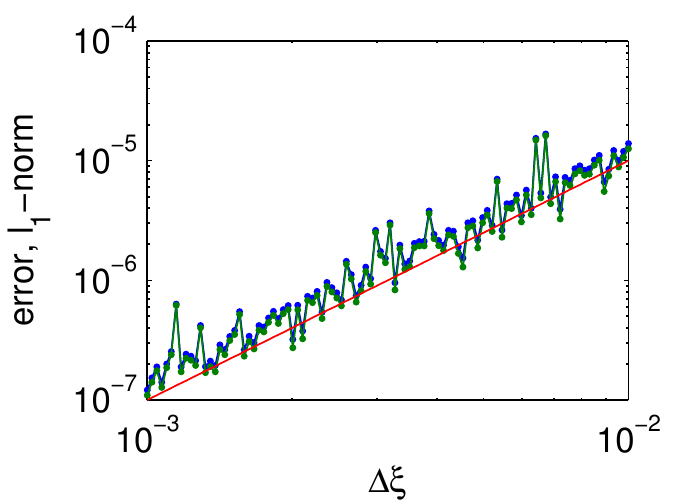}
  \caption{Convergence of boundary integral method
    \eqref{eq:BIE_u_seg}, for $u$ and $v$, with second order given by
    solid line. Left: random sample set of observation points $\x_0$
    away from the interface. Right: Observation points on the
    interface, i.e. singular integrands.}
  \label{fig:bie_conv}
\end{figure}

The quadrature method described is trapezoidal, but with one interval
of unequal length. In Figure \ref{fig:bie_conv} we establish second
order convergence in $\infty$-norm away from the interface and second
order convergence in $l_1$-norm on the interface, as the interface is
refined. Here we have taken the interface circular, so that, by
symmetry, the exact solution is $\U = 0\,\, \forall \x_0$. Away from
the interface we randomize a set of observation points, and on the
interface we take as observation points the entire interface.

%% file: ns.tex
\section{Incompressible Navier Stokes flow computations}
\label{sec:NS}

We now demonstrate that our interface tracking method is applicable to
the well established application of two-phase Navier-Stokes flow and
that good results are obtained. There is not much here which is
specific to our method. Indeed, this is an important point -- our
interface tracking method works well with established components in
multi-phase flow computations.

\subsection{Overview}
The task of solving a multi-phase Navier-Stokes flow system is
complicated for several reasons. At a basic level, solving the
Navier-Stokes equations themselves is a challenging problem on its
own. The incompressible Navier Stokes equations are given by
\begin{align}
  \rho \left(\pd{\U}{t} + (\U\cdot\nabla)\U\right)  = -\nabla p + 
  \mu \nabla^2 \U + \F \label{eq:ns_momentum}\\
  \nabla \cdot \U = 0\label{eq:ns_continuity}\\
  + (\text{Boundary conditions}). \label{eq:ns_bc}
\end{align}
where $\mu$ denotes viscosity and $\rho$ density. We will, for the
remainder of this paper, consider the momentum equation in
non-dimensional form:
\begin{align}
  \pd{\U}{t} + (\U\cdot\nabla)\U  = - \nabla p + 
  \frac{1}{\RE} \nabla^2 \U + \F, \label{eq:ns_momentum_nd}
\end{align}
with $\RE := \frac{U L}{\mu}$, the Reynolds number.

The two main difficulties here are that the first (i.e. momentum)
equation is non-linear, and that the space of admissible solutions is
restricted (by the second equation) to be divergence free in the
entire domain. We refer the reader to the excellent textbook by
Acheson \cite{Acheson1990}. Several classes of numerical methods exist
for Navier-Stokes, based on finite volume (FV), finite difference (FD)
and finite element (FEM), as well as spectral methods. These often
have high accuracy orders with respect to the spatial
discretization. There are, however, much fewer methods that are fully
second order accurate also in time. In the fluid mechanics community,
this is generally regarded as of little consequence, since there are
often physical time scales in the flow that are small. In this work we
use a fairly basic pressure correction FD method, as presented in
Section \ref{sec:NS_method}.

When a dynamic interface is added, e.g. the case of two immiscible
liquids, the picture becomes more complicated still. The physical
parameters of the flow may be different in the different fluids, and a
two way coupling is introduced between the interface and the flow
field. That is, the fluid acts upon the free surface which
simultaneously acts upon the fluid. From a mathematical point of view
it is clear that the equations of the two (or more) fluids and the
dynamics of the free surface constitute one large, coupled and highly
non-linear system. Furthermore, application specific challenges arise
frequently. In the ubiquitous case of a bubble in a bulk fluid the
most widely used formulation introduces a singular source term in the
momentum equation \eqref{eq:ns_momentum} to mediate forces on the
fluid from the interface, e.g. surface tension. The lack of regularity
in the velocity and pressure solutions to the Navier Stokes equations
presents significant numerical challenges. Here, we shall use the
immersed boundary (IBM) and immersed interface (IIM) methods as
presented in the following sections.

\subsection{Numerical method for Navier Stokes} \label{sec:NS_method}
The numerical method to treat the Navier Stokes equations
\eqref{eq:ns_momentum}-\eqref{eq:ns_bc} is a finite difference method
on a staggered (MAC) grid, which employs so called \emph{Chorin
  splitting}. It is presented in detail in the textbook by Strang
\cite[Ch. 6.7]{Strang2007}, and a basic implementation due to Seibold
is available on-line \cite{SeiboldCode2007}. We summarize the
time-discrete method here:

Compute a first intermediate velocity
field, $\U^{*}$, by treating the non-linear term explicitly,
\begin{equation} 
  \label{eq:ns_sd_I}
  \frac{\U^{*}-\U^n}{\dt} = (\U^n\cdot\nabla)\U^n + \F.
\end{equation}

To avoid a severe time-step restriction the
diffusive term is treated implicitly and a second intermediate is
obtained via
\begin{equation}
  \label{eq:ns_sd_II}
  \frac{\U^{**}-\U^*}{\dt} = \frac{1}{\RE}\Delta \U^{**}.
\end{equation}
This, together with a discrete approximation for the spatial
derivatives, constitute a linear system. Several appropriate solvers
exist for this system depending on the memory constraints and the
boundary conditions applied. The final result is obtained by applying a
correction step,
\begin{equation}
  \label{eq:ns_sd_IIIa}
  \frac{\U^{n+1}-\U^{**}}{\dt} = -\nabla p^{n+1},
\end{equation}
based on the pressure, which is obtained by solving a Poisson equation,
\begin{equation}
  \label{eq:ns_sd_IIIb}
  -\Delta p^{n+1} = -\frac{1}{\dt}\nabla\cdot\U^{**}.
\end{equation}
We use central difference approximations for the spatial derivatives,
so that $\nabla$ is replaced by the standard discrete nabla operator
on a staggered grid $\nabla_h$. However, when applied to the
convection step a moderate upwind bias is applied so as to not
introduce a numerical instability (see \cite[Ch. 6.7]{Strang2007}).

Several remarks are in order. The pressure correction strategy is by
no means the only one available for the task of decoupling the
momentum equation from the incompressibility constraint. Another
popular approach is the $\theta$-splitting method (see e.g. Glowinski
and Pironneau \cite{GlowinskiPironneau1992}). A large body of work
exist on the topic of pressure correction methods for Navier
Stokes. The approach here dates back to Chorin \cite{Chorin1968} and
may be seen as the most elementary method. Several more recent methods
were presented and evaluated in a recent survey by Guermond
et. al. \cite{Guermond2006}. The class of methods that Guermond refers
to as \emph{consistent splitting methods} can be seen to be more
accurate than Chorin splitting. We also remind the reader that
pressure correction methods do not necessarily have the property that
both the non-divergence constraint and the prescribed boundary
conditions are accurately satisfied simultaneously (due, at least in
part, to the lack of a consistent numerical boundary condition for the
pressure). Shedding light on these interesting topics is beyond the
scope of this work, and we also content ourselves with using a first
order accurate time discretization. A final remark concerns the
efficiency of the method above: the linear solves in the diffusion and
pressure correction steps involve constant system matrices. These
matrices need only be factorized once, leaving just the back- and
forward substitution steps to be performed in the time loop. The major
constraint here is clearly the memory requirements of the factorized
matrices, which may be significantly less sparse, but the efficiency
obtained is of great practical value.

The flow solver couples to the interface tracking method previously
presented, using Strang splitting \eqref{eq:strang}.

\subsection{Interface/flow coupling: immersed boundary formulation}
As a basic case, let $\Gamma$ be a closed interface separating two
liquids with equal density and viscosity. Two main approaches exist
for coupling the motion of a dynamic interface with a incompressible,
viscous bulk flow (described by the Navier Stokes equations
above). First, separated grids may be generated for the separated flow
regions -- so called body fitted methods. The second class of methods
follows the work of Peskin and collaborators \cite{Peskin1977,
  Peskin1980, Peskin1993, Peskin2002} on the immersed boundary (IB)
method, where one views the separated flow regions as one system and a
suitable force density is added to mediate the action of the interface
on the fluid system. This method dates back to the 1970's but still
sees a remarkable amount of use. The 2002 survey paper by Peskin may
be of particular interest to some readers \cite{Peskin2002}.

Here, we shall use the IB formulation, with surface tension driving
the flow internally. It is well known that the surface tension force
as a function of Lagrangian variables on the interface can be
formulated as
\begin{align}
  \f(s,t) = \sigma\kappa(s,t) \nhat(s,t),
\end{align}
where $\sigma\in \R^+$ is the surface tension coefficient, $\kappa$ is
the interface curvature and $\nhat$ is the unit inward normal. Recall
that the capillary number, $\CA$, is defined as $\CA:= \frac{U
  \mu}{\sigma}$. We may then write the non-dimensionalized force that
will enter the momentum equation \eqref{eq:ns_momentum_nd} in Eulerian
variables as
\begin{align}
  \F(\x,t) &= \frac{1}{\RE\,\CA} \int \kappa(s,t) \nhat(s,t)\delta_\Gamma\d s, 
  \label{eq:ns_force_spread}
\end{align}
where $\delta_\Gamma$ is a Dirac measure on the interface,
i.e. $\delta_\Gamma(\x) = \delta(\x - \Gamma(s))$ and $\delta(\x)$ is
the composition of one-dimensional delta functions, $\delta(\x) =
\delta(x_1)\delta(x_2)$. Transformations from Lagrangian to Eulerian
variables in this way are at the heart of the IB method and much
related work. It is also appropriate to precisely define how the
convective field in the interface advection ODE
\eqref{eq:interface_ode} arises:
\begin{align}
  \U(\Gamma,t) := \int_\Omega \U(\x,t) \delta_\Gamma \d\x.
  \label{eq:ns_vel_eval}
\end{align}

In order to complete the formulation of the discrete immersed boundary
method one needs to introduce discrete delta function
approximations. We refer the reader to a study by Tornberg and
Engquist on the treatment of singular source terms
\cite{Tornberg2004a}. We use the piecewise cubic delta function
approximation introduced therein,
\begin{align}
  \delta_{2h} (x) = 
  \begin{cases}
    \frac{1}{h} \phi(x/h), & \quad |x|\leq 2h \\
    0,&  \quad |x| > 2h
  \end{cases},
\end{align}
with
\begin{align}
  \phi(r) = 
  \begin{cases}
    1 - \frac{1}{2}|r| - |r|^2 + \frac{1}{2}|r|^3, & 0 \leq r \leq 1 \\
    1 - \frac{11}{6}|r| + |r|^2 - \frac{1}{6}|r|^3, & 1 < r \leq 2. 
  \end{cases}
\end{align}
This delta function has narrow support on the grid and possesses a
larger number of discrete moments than other commonly used
variants. There are other good choices for the discrete delta function
approximation. However, there is a generally held view, concerning
stability, that the same approximation should be used in the
``spreading'' step \eqref{eq:ns_force_spread} as in the evaluation
step \eqref{eq:ns_vel_eval}.

\subsection{Immersed interface method}
\label{sec:NS_iim}
The singular source term introduced by the IBM gives rise to a jump
discontinuity in the pressure, which is well understood physically:
the pressure gradient balances the interfacial forces. This jump is
\begin{align}
  \jmp{p}(s,t) = \f(s,t)\cdot\nhat(s,t). 
\end{align}
That is, the pressure jump equals the magnitude of the interfacial
force in the normal direction, which is simply $\jmp{p}(s,t) =
\sigma\kappa(s,t)$ in our case. Additional jump conditions need to be
considered in settings where there are e.g. tangential stresses on the
interface.

The immersed interface method (IIM), introduced and developed by
LeVeque and collaborators in a series of papers \cite{Leveque1994,
  Leveque1997, Lee2003}, is in essence a evolution of the IBM where
the jump conditions are corrected for in the numerical method rather
than taken into the momentum equation as a source term. The textbook
by Li and Ito \cite{Li2006}, and a recent paper by Le et. al
\cite{Le2006} may also be of interest. Derivations of the correction
terms are given in the works cited above, and we humbly restate the
relevant results here.

Going back to the NS method \eqref{eq:ns_sd_I}-\eqref{eq:ns_sd_IIIb},
drop the source term, $\F$ completely from the convective step,
\eqref{eq:ns_sd_I}. The diffusive step \eqref{eq:ns_sd_II} is
unaltered, and the pressure correction step \eqref{eq:ns_sd_IIIa} and
\eqref{eq:ns_sd_IIIb} we replace by 

\begin{equation}
  \label{eq:ns_sd_iim}
  \left\{
    \begin{array}{l}
      \frac{\U^{n+1}-\U^{**}}{\dt} = -\nabla p^{n+1} + \mathbf{B}\\
      -\Delta p^{n+1} = -\frac{1}{\dt}\nabla\cdot\U^{**} + C.
    \end{array}
  \right.
\end{equation}

The correction terms, $\mathbf{B} = (B^1, B^2)$ and $C$, only have
meaningful definitions as discrete functions. Recall that we are
working on a staggered (MAC) grid. It turns out to be natural to
evaluate the correction term $B^1$ at intersections with horizontal
grid lines, so that $B^1_{i-1/2,j} = \jmp{p}/\Delta x$ if the
interface cuts the grid between points $(i,j)$ and $(i-1,j)$, and zero
otherwise. Correspondingly, the correction on the $y$-grid is given by
$B^2_{i,j-1/2} = \jmp{p}/\Delta y$ if the interface cuts the vertical
grid lines between $(i,j)$ and $(i,j-1)$, and zero otherwise. Finally,
\begin{align}
  B^1_{ij} = \frac{1}{2}\left( B^1_{i-1/2,j} + B^1_{i+1/2,j} \right), \\
  B^2_{ij} = \frac{1}{2}\left( B^1_{i,j-1/2} + B^1_{i,j+1/2} \right),\\
  C_{ij} = (\nabla_h \cdot \mathbf{B})_{ij}.
\end{align}
This concludes the formulation of the discrete immersed interface
method.

\subsubsection{IBM vs. IIM: Accuracy} 
Due to the smearing of the interface introduced by the regularized
delta function, the IBM will at most attain first order accuracy for
problems with non-smooth solutions.  The IIM has been shown to produce
second order accurate results.

\subsubsection{IBM vs. IIM: Stability} 
Interestingly, there has been much recent work on the stability of the
IBM, despite the many years since it was introduced. Peskin himself
emphasizes stability as one of the outstanding challenges for the IBM
in his Acta Numerica paper \cite{Peskin2002}. Newren
et. al. \cite{Newren2007} appear to have resolved this issue by
providing a general stability theory for the IBM applied to Stokes
flow. They show unconditional stability for a Crank-Nicholson method
and backward Euler method. Following that paper, Newren
et. al. \cite{Newren2008} discuss practical solver strategies for
semi- and fully implicit immersed boundary methods for Navier
Stokes. They substantiate the widely held view that fully explicit
methods are competitive, despite the short time steps, due to the high
computational costs associated with the iterative methods needed in
implicit IBM formulations. In another recent paper, Hou and Shi
\cite{Hou2008} also obtain unconditionally stable discretizations for
the Stokes case. Based on earlier work by Hou, Lowengrub and Shelley
\cite{Hou1994} and their unconditionally stable method, they go to
great lengths to derive an efficient semi-implicit method with good
stability properties. While this appears to be the most efficient
method known at present, it relies heavily on spectral properties of
the spatial discretization and, hence, is only applicable in certain
cases.

Newren et. al. point out two conditions as sufficient to prove
stability of immersed boundary methods: First, the spreading and
interpolation must be adjoint operators. Second, the velocity field
needs to be discretely divergence free. Hou and Shi reiterate this
conclusion. Hou and Shi also make a separate point about the need to
alleviate the stiffness in the interface/flow coupling, and go about
it by using the \emph{arc-length and tangent angle} formulation
(cf. \cite{Hou1994, Hou2008}).

Little is known to date about the stability properties of immersed
interface methods. There are several vague comments on stability in
the IIM literature \cite{Leveque1997, Lee2003, Le2006, Li2006}. The
authors seem to agree that the explicit methods suffer from a severe
stiffness, which motivates implicit methods. Going back to the
sufficient conditions for stability of IB methods, it is clear that
there is no adjointness property applicable in the IIM case (since
there is no spreading operator). However, one could reasonably
expect the non-divergence condition to be better handled by the IIM:
In the IBM, as the pressure jump gets bigger the smeared pressure
becomes under-resolved numerically and oscillatory near the
interface. This would lead to a less accurate pressure correction,
and hence a failure to be accurately divergence free in the
velocity. The pressure solve is more accurate in the IIM. Further,
the idea of formulating the interface problem in terms of arc-length
and tangent angle has not been applied to the IIM as far as we know.


\subsection{Numerical results: Navier-Stokes}

\subsubsection{Shear flow} \label{sec:ns_shear} 

\begin{figure}
  
  \includegraphics[width=.79\columnwidth]{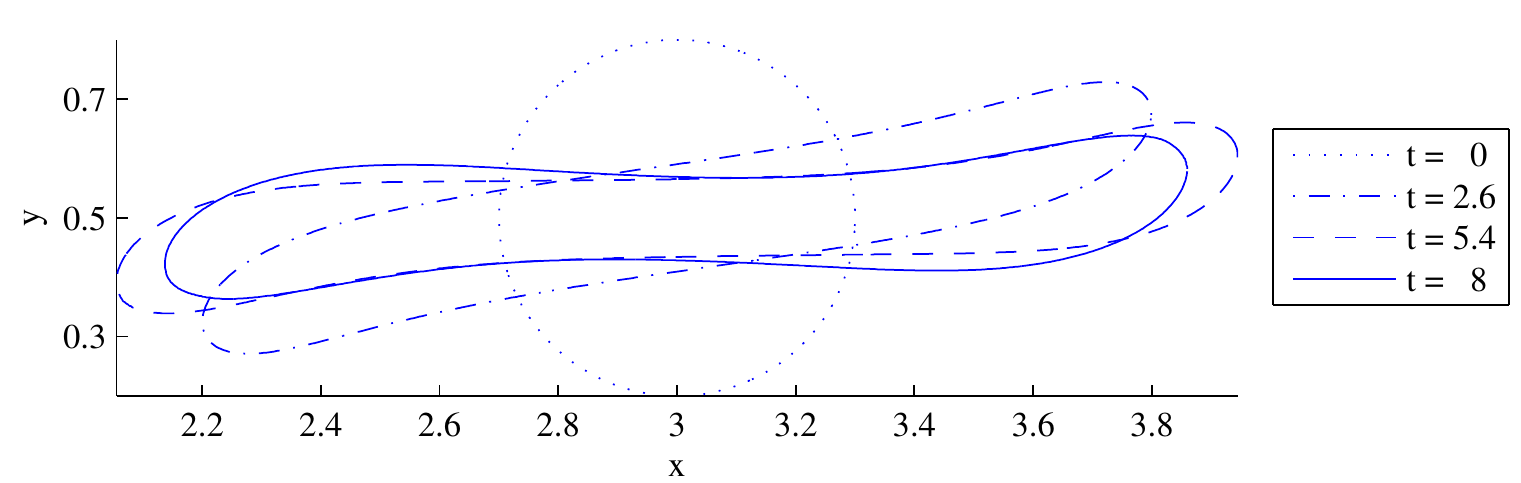}
  \includegraphics[width=.65\columnwidth]{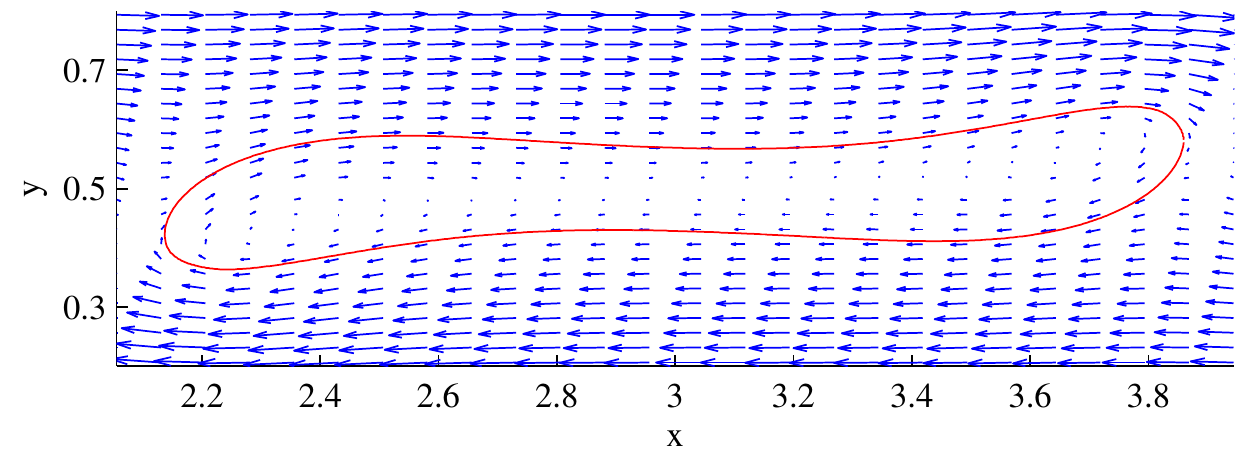}
  
\includegraphics[width=.79\columnwidth]{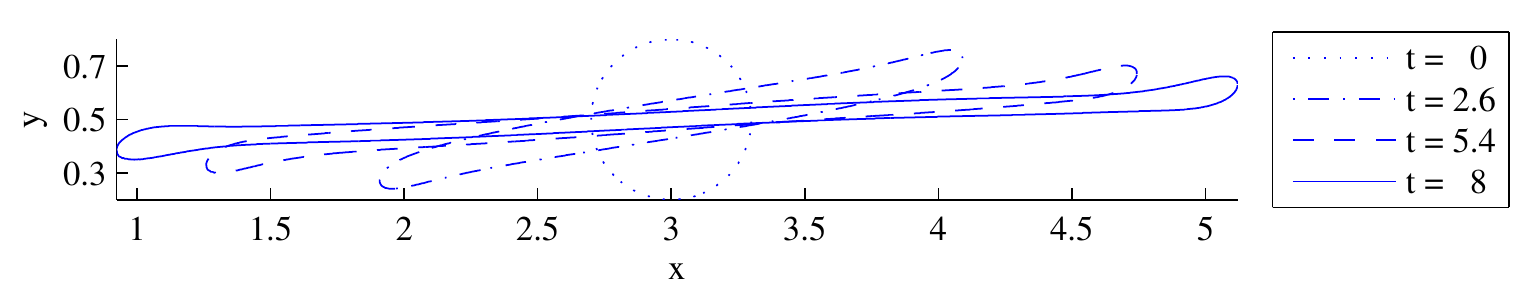}
  \includegraphics[width=.65\columnwidth]{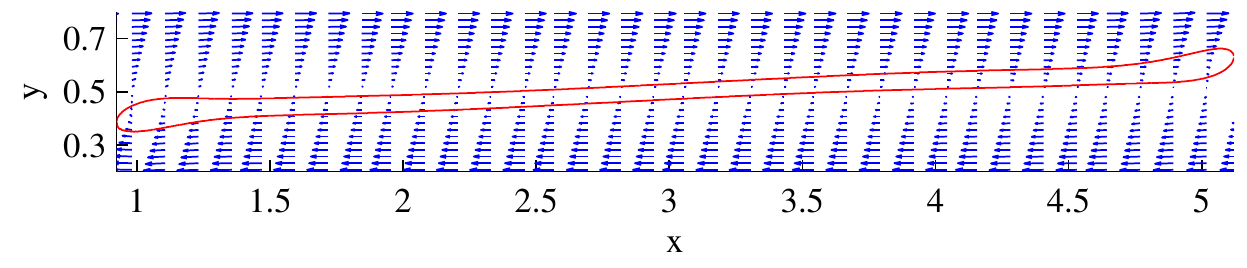} 
  \caption{Interface location at various times, and flow field at
    $t=8$. Top pair: $\RE=1, \CA=1$. Below critical $\CA$, the
    interface reaches a steady state. Bottom pair: $\RE = 1,
    \CA=2$. Above critical $\CA$, no steady interface is obtained.}
  \label{fig:ns_shear_1}
\end{figure}

We illustrate the NS/IBM solver by considering a drop in shear
flow. The shear velocity is imposed as a Dirichlet boundary condition
on the top and bottom boundaries, ramped up via
\begin{align*}
  u(1,t) =
  \begin{cases}
    \frac{u_\infty}{2} \left(1-\cos\left(\frac{\pi
          t}{t_\infty}\right) \right), & t\leq t_\infty\\
    u_\infty & t > t_\infty
  \end{cases}
\end{align*}
for some ramp-up time $t_\infty$ and shear velocity $u_\infty$. We let
$u(0,t) = -u(1,t)$.  In the $x$-direction we impose periodic boundary
conditions. In Figure \ref{fig:ns_shear_1} we present computational
results with different coefficients of surface tension. Here we used
$u_\infty=1$, $t_\infty = 1/5$ and 80 grid-points in the
$y$-direction. These results could equally well have been obtained
with the immersed interface method given above. Throughout this
section we use the channel height, $L_y=1$, when determining the
Reynolds number.

\subsubsection{Convergence} \label{sec:ns_conv} 

The method used here is not more than first order accurate, as
discussed prior. To verify this we ran numerical convergence tests
using a shear flow setup. Here we use an initially circular drop
with radius $r = 1/3$ centered in a unit square domain. The shear
boundary conditions are ramped up as above, with $t_\infty=1/5$ and we
compute until $t=1/4$. With $\RE = \CA = 1$ we get, at that time, that
the interface is moderately deformed, but not close to a steady state.

In Tables \ref{tab:num_conv_shear_ibm} and
\ref{tab:num_conv_shear_iim} we give convergence results for the
immersed boundary and immersed interface methods respectively, as the
grid is refined. The ratio $h/\dt$ is kept fixed, so several hundred
time steps are taken on the finest mesh. These results indicate the
expected overall first order convergence for both methods.

Here we let $u_i$ be a sequence of computed solutions, and assume the
error with respect to some unknown exact solution $\tilde{u}$ behaves
like $ \|u_i -\tilde{u}\| \approx O(h_i^p) + O(\dt^p)$. We define
$\Delta^i := \| u_{i} -u_{i-1} \|$ for various norms and have
estimates of the convergence order $p = \log_2(\Delta^{i+1}/\Delta^i)$
as the grid is refined by a factor 2. To see a benefit in terms of
accuracy from using the immersed interface method, a second order
accurate time stepping method for the flow solver would have to be
implemented.

\begin{table}
  \centering
  \begin{tabular}{l|ll|ll}
    $N$ & $\Delta_u$ & $p_u$ &$\Delta_v$ & $p_v$\\
    \hline
    44 & & & &  \\ 
    88 & 0.011282 & & 0.000421 & \\ 
    176 & 0.005611 & 1.007641 &  0.000202 & 1.056737 \\ 
    352 & 0.002873 & 0.965510 &  0.000098 & 1.042040 \\
  \end{tabular}
  ~\\ ~ \\[.5 cm]
  \begin{tabular}{l|ll|ll}
    $N$ & $\Delta_u$ & $p_u$ &$\Delta_v$ & $p_v$ \\
    \hline
    44 & & & & \\ 
    88 & 0.011329 & & 0.000540 & \\ 
    176 & 0.005636 & 1.007214 &0.000255 & 1.082531\\ 
    352 & 0.002881 & 0.968202 &0.000124 & 1.039785\\
  \end{tabular}
  ~\\ ~ \\[.5 cm]
  \begin{tabular}{l|ll|ll}
    $N$ & $\Delta_u$ & $p_u$ &$\Delta_v$ & $p_v$\\
    \hline
    44 & & & & \\ 
    88 & 0.013639 & & 0.002467 &  \\ 
    176 & 0.006866 & 0.990110 & 0.001237 & 0.995364\\ 
    352 & 0.003483 & 0.979443 & 0.000658 & 0.912272\\
  \end{tabular}
  \caption{ NS/IBM grid refinement study, shear flow, 
    Top: $l_1$, Middle: $l_2$, Bottom: $l_\infty$. See Section 
    \ref{sec:ns_conv}}
  \label{tab:num_conv_shear_ibm}
\end{table}

\begin{table}
  \centering
  \begin{tabular}{l|ll|ll}
    $N$ & $\Delta_u$ & $p_u$ &$\Delta_v$ & $p_v$\\
    \hline
    44 & & & &  \\ 
    88 & 0.011935 & & 0.000493 & \\ 
    176 & 0.005223 & 1.192320 & 0.000282 & 0.807506 \\ 
    352 & 0.002829 & 0.884727 & 0.000107 & 1.391809 \\
  \end{tabular}
  ~\\ ~ \\[.5 cm]
  \begin{tabular}{l|ll|ll}
    $N$ & $\Delta_u$ & $p_u$ &$\Delta_v$ & $p_v$ \\
    \hline
    44 & & & & \\ 
    88 & 0.011969 & & 0.000645 & \\ 
    176 & 0.005274 & 1.182247 & 0.000378 & 0.768790 \\ 
    352 & 0.002841 & 0.892364 & 0.000144 & 1.391774 \\
  \end{tabular}
  ~\\ ~ \\[.5 cm]
  \begin{tabular}{l|ll|ll}
    $N$ & $\Delta_u$ & $p_u$ &$\Delta_v$ & $p_v$\\
    \hline
    44 & & & & \\ 
    88 & 0.015243 & & 0.004002 & \\ 
    176 & 0.008466 & 0.848441 & 0.001872 & 1.096254 \\ 
    352 & 0.004081 & 1.052766 & 0.000742 & 1.335409 \\ 
  \end{tabular}
  \caption{ NS/IIM grid refinement study, shear flow, 
    Top: $l_1$, Middle: $l_2$, Bottom: $l_\infty$. See Section 
    \ref{sec:ns_conv}}
  \label{tab:num_conv_shear_iim}
\end{table}

\subsubsection{Brief comparative study} \label{sec:comp}

Whereas a full comparison between the immersed interface- and
boundary method including solution methods would be valuable, it is
beyond the scope of this work. The space of parameters,
e.g. discretization and solution methods, capillary- and Reynolds
numbers, is \emph{very} large. Still, we think some brief comparisons are in
order.

\begin{figure}
  \centering
  \includegraphics{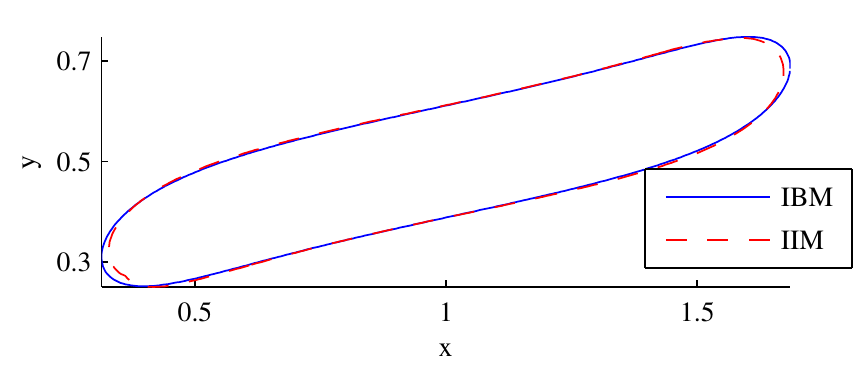} 
  \caption{Interface in shear flow, at time $T=2$, with both IIM and IBM.}
  \label{fig:ns_comp_001}
\end{figure}

First, for the simple shear flow setup described above we compare the
results obtained with the immersed interface- and boundary methods
with explicit time stepping. As can be seen in Figure
\ref{fig:ns_comp_001}, there is a small but visible discrepancy. Here,
the domain was $[0\,\, 2]\times[0\,\, 1]$, with a grid with
$h_x=h_y=1/100$ and the interface was discretized with
$\dxi=1/200$. Again, the physical parameters were $\RE=\CA=1$. In this
test we noted that the IIM was more prone to going unstable, as
expected. We had to take $\Delta t = 1/300$ with the IBM and $\Delta
t= 1/3000$ with the IIM, to get stable computations.

For a more challenging case, e.g. with $\CA$ several orders below
unity, we expect to see the implicit (CN) method stable for larger
time steps than the explicit method. However, this has not been
evident to us -- at least not to the extent that the overall
computational cost of a run becomes smaller with an implicit
method. That implicit methods fail to be competitive for the IBM is in
line with results by Newren et. al. \cite{Newren2008} and others. We
have to conclude with noting that more research is need to illuminate
the efficient deployment of immersed boundary- and interface methods.

\begin{figure}
  \centering
  \includegraphics{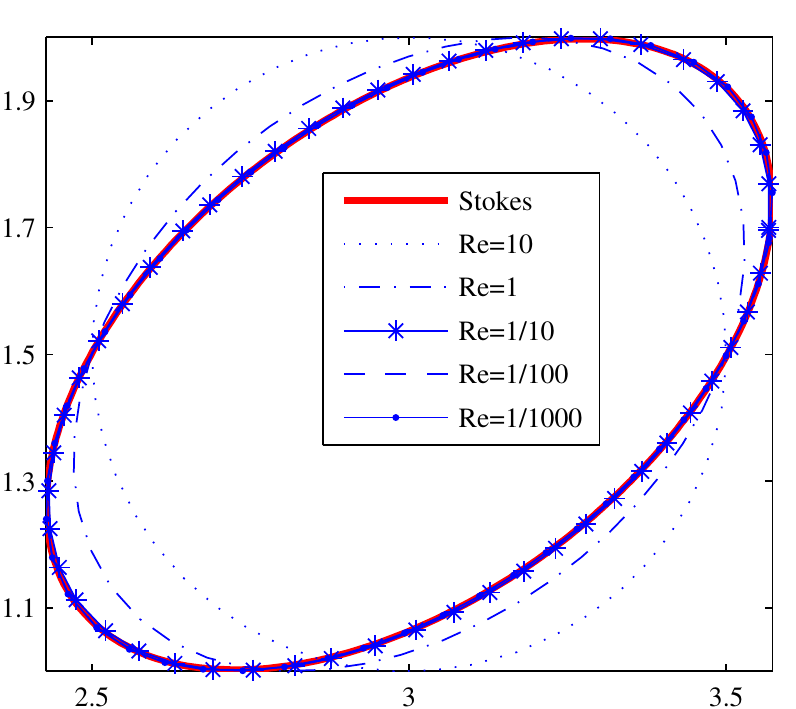} 
  \includegraphics{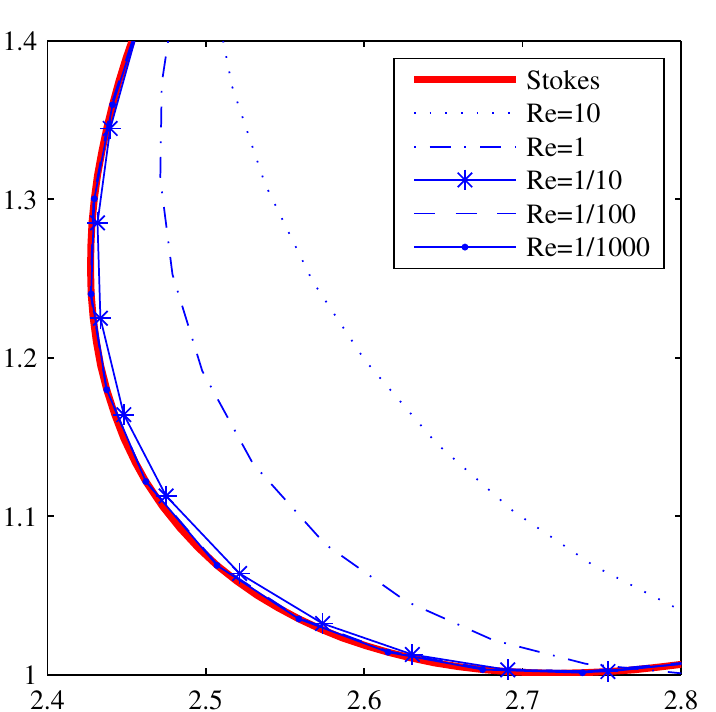} 
  \caption{Interface in shear flow, at time $T=1$, and $\CA = 5$. We
    illustrate that as $\RE\rightarrow 0$, the Navier-Stokes (IBM)
    method converges to the results from the boundary integral Stokes
    method (cf. Section \ref{sec:BIE_Stokes}). Here the interface is
    initially circular, with diameter 1, and at location $(3,3/2)$, in
    a domain $[0, 6]\times[0, 3]$ and shear boundary conditions
    discussed in Section \ref{sec:ns_shear}. Note that the $T=1$ is
    before a steady state is reached. Bottom: close-up from above.}
  \label{fig:stokes_shear_001}
\end{figure}

In Figure \ref{fig:stokes_shear_001} we consider a shear flow case
where the Reynolds number goes to zero, i.e. the limit of Stokes
flow. Here we compare the results from the Navier-Stokes (IBM) method
with the boundary integral Stokes method (Section
\ref{sec:BIE_Stokes}) and find excellent agreement as $\RE\rightarrow
0$.

